%% file: main.tex
\documentclass[acmsmall,
  bookmarksnumbered,
  unicode,
  colorlinks=true,
  citecolor=blue,
  linkcolor=black,
  urlcolor=blue
]{acmart}

\AtBeginDocument{%
  }

\setcopyright{none}

\usepackage[ruled,linesnumbered,noend]{algorithm2e}
\usepackage{algpseudocode}
\usepackage{subcaption}
\usepackage{bbding}
\usepackage{enumitem}
\usepackage{makecell}
\usepackage{graphicx}
\usepackage{grffile}
\usepackage{cases}
\usepackage{empheq}
\usepackage{booktabs}

\usepackage{amssymb} 
\usepackage{multirow} 
\usepackage{bm}
\usepackage{pifont}
\usepackage{diagbox}

\newcommand{\ourmethod}{\textsc{FAVOR}\xspace}
\newcommand{\ouralg}{\text{FAVOR}\xspace}

\newcommand{\ie}{\textit{i}.\textit{e}.}
\newcommand{\eg}{\textit{e}.\textit{g}.}

\newcommand{\zl}[1]{{\color{black}#1}}

\begin{document}

\title{FAVOR: Efficient Filter-Agnostic Vector ANNS Based on Selectivity-Aware Exclusion Distances}

\author{Junjie Song}
\orcid{0009-0001-3891-783X}
\affiliation{%
  \institution{Huazhong University of Science and Technology}
  \country{China}}
\email{d202381493@hust.edu.cn}


\author{Yu Liu}
\authornote{Corresponding author.}
\orcid{0000-0002-1964-9278}
\affiliation{%
  \institution{Huazhong University of Science and Technology}
  \country{China}}
\email{liu_yu@hust.edu.cn}
\affiliation{%
  \institution{National University of Singapore}
  \country{Singapore}}

\author{Guoyu Hu}
\orcid{0009-0005-9463-2045}
\affiliation{%
  \institution{National University of Singapore}
  \country{Singapore}}
\email{guoyu.hu@u.nus.edu}

\author{Zhongle Xie}
\orcid{0000-0002-2924-6974}
\affiliation{%
  \institution{Zhejiang University}
  \country{China}}
\email{xiezl@zju.edu.cn}

\author{Ming Yang}
\orcid{0009-0001-6925-6309}
\affiliation{%
  \institution{Wuhan Technical University}
  \country{China}}
\affiliation{%
  \institution{National University of Singapore}
  \country{Singapore}}
\email{mingyang@wtc.edu.cn}

\author{Beng Chin Ooi}
\orcid{0000-0003-4446-1100}
\affiliation{%
  \institution{Zhejiang University}
  \country{China}}
\email{ooibc@zju.edu.cn}

\author{Ke Zhou}
\orcid{0000-0002-2161-8796}
\affiliation{%
  \institution{Huazhong University of Science and Technology}
  \country{China}}
\email{zhke@hust.edu.cn}

\renewcommand{\shortauthors}{Junjie Song et al.}

\begin{abstract}
Modern retrieval systems increasingly require integrating approximate nearest neighbor search (ANNS) with complex attribute filtering to handle hybrid queries in applications such as recommendation systems and retrieval-augmented generation (RAG). 
While HNSW-based inline-filtering methods show promise, existing approaches struggle to deliver high throughput under low-selectivity scenarios while balancing search efficiency, filtering generality, and index connectivity. 
To address these challenges, we propose \ourmethod, an efficient filter-agnostic vector ANNS that supports arbitrary filtering conditions while maintaining stable performance across varying selectivity levels.
\ourmethod
introduces three novel features:
(1) an integrated architecture that unifies selectivity estimation and filtered ANNS execution, providing a cohesive solution for hybrid vector–attribute queries; 
(2) a HNSW-based inline-filtering algorithm that introduces an exclusion distance mechanism to dynamically reshape the vector distance distribution, pushing non-target vectors away from the query while promoting valid candidates toward the query, thus improving search efficiency without compromising generality or graph connectivity; and (3) a selectivity-driven search selector that estimates query selectivity and dynamically routes queries between a pre-filtering brute-force algorithm for low-selectivity cases and an optimized HNSW-based search algorithm for other scenarios, ensuring consistent performance.
Extensive experiments on real-world datasets demonstrate that \ourmethod achieves a 1.3–5$\times$ higher QPS at $Recall@10 = 95\%$ compared to state-of-the-art methods for arbitrary filtering conditions, while maintaining competitive performance even against tailored solutions in some filtering conditions.
\end{abstract}

\ccsdesc[500]{Information systems~Top-k retrieval in databases}
\ccsdesc[500]{Information systems~Similarity measures}
\ccsdesc[300]{Information systems~Query processing}
\ccsdesc[300]{Information systems~Data structures}

\keywords{Approximate nearest neighbor search, Vector databases, AIxDB}
%


\maketitle

\input{Introduction}
\input{Preliminary}

\input{FAVOR}
\input{Selector}
\input{FAVORsearch}
\input{evaluation}
\input{relatedwork}
\input{conclusion}
\input{Acknowledgments}


\bibliographystyle{ACM-Reference-Format}
\bibliography{references}


\appendix

\end{document}

%% file: Introduction.tex
\section{INTRODUCTION}
\label{sec:intro}

Recent advances in deep learning have enabled the encoding of rich semantic information from unstructured data into high-dimensional vectors~\cite{doc2vec, graph2vec_1}. 
These embeddings support semantic retrieval through nearest neighbor search in vector space. However, the paradigm often falls short in real-world scenarios where queries involve both semantic relevance and structured constraints. For example, when users search for videos on YouTube, they may provide a textual description while also specifying filters such as upload date or video duration. Such hybrid queries, combining semantic similarity with attribute-based filtering, are increasingly common in recommendation systems, search engines, and retrieval-augmented generation (RAG) applications~\cite{rag1, rag2}.

Approximate nearest neighbor search (ANNS) utilizing graph-based indexes has gained widespread adoption due to its efficient search and superior retrieval quality~\cite{milvus, graph-anns-bench}. 
Within this category, Hierarchical Navigable Small World (HNSW) has emerged as the predominant methodology, distinguished by its capacity for incremental construction and 
\zl{
relatively few tuning parameters compared with alternative methods~\cite{HNSW}.
}
Yet, integrating HNSW with attribute filtering remains challenging because the subset of vectors satisfying filtering conditions often forms a sparsely connected subgraph. 
Specifically, there are three primary challenges in this domain. 

\begin{itemize}[noitemsep,topsep=0pt,parsep=0pt,partopsep=0pt,leftmargin=*]
    \item \textbf{Challenge 1 (C1): Supporting arbitrary filtering conditions.} 
    The need for attribute-tailored indexes to support diverse filtering types creates a fundamental tension between query efficiency and the practical costs of index construction and maintenance. 
    For example, tailored approaches~\cite{ung, SIEVE, SeRF} redesign the index structure to explicitly connect vectors satisfying specific attributes, thereby avoiding unnecessary similarity computations.
    However, these tailored indexes lack scalability and flexibility, as they cannot efficiently generalize to new filtering conditions.
    \item \textbf{Challenge 2 (C2): Balancing search efficiency and recall.} Preserving graph connectivity requires expensive similarity computations on vectors that violate filtering conditions, making it exceptionally challenging to optimize for both efficiency and recall. 
    Filter-agnostic methods~\cite{ACORN, HNSW} address this by modifying the search process while retaining the index structure, but they often fail to balance optimal path exploration and computational efficiency, leading to substantial performance gaps compared with tailored approaches.
    \item \textbf{Challenge 3 (C3): Maintaining stable performance under varying selectivity.} The inherent fluctuation in query selectivity poses a significant risk to search performance, which can degrade sharply with low-selectivity queries.  
    For instance, under low selectivity, ACORN~\cite{ACORN} suffers from connectivity issues, whereas result-set filtering~\cite{HNSW} struggles to terminate efficiently due to the cost of finding sufficient filtering-valid vectors closer to the query vector than those in the candidate set.  
\end{itemize}

To tackle these challenges, we propose \ourmethod, an efficient \underline{F}ilter-\underline{A}gnostic \underline{V}ect\underline{OR} (\ourmethod) ANNS. 
\ourmethod seamlessly integrates the estimation of query selectivity with two distinct search algorithms, each adaptive for different selectivity levels. It includes a selectivity-driven search selector, an \ourmethod search algorithm, and a pre-filtering brute-first algorithm.
By employing a standard proximity graph and a unified query processing workflow, \ourmethod supports arbitrary filtering conditions without incurring the construction and maintenance overhead of tailored indexes (addressing \textbf{C1}). At the heart of our search algorithm is an exclusion distance mechanism. By dynamically adjusting vector distributions with a selectivity-aware exclusion distance based on filtering compliance and enabling filtering-valid vectors to be closer to the query vector than their neighbors, it steers the search towards valid results while preserving connectivity, effectively balancing the connectivity and efficiency (addressing \textbf{C2}). The integrated selectivity-driven search selector determines the optimal choice between a pre-filtering and \ourmethod search algorithm for each query, ensuring stable performance across varying selectivity levels (addressing \textbf{C3}). Together, \ourmethod enables selectivity-aware, filter-agnostic ANNS with performance on par with or superior to tailored methods.  

Our contributions are summarized as follows:
\begin{itemize}[noitemsep,topsep=0pt,parsep=0pt,partopsep=0pt,leftmargin=*]
    \item 
    We propose \ourmethod, a filter-agnostic vector ANNS scheme, which integrates a selectivity-driven search selector and two complementary search algorithms.
    It achieves stable query performance across the full selectivity spectrum, leveraging a pre-filtering brute-force algorithm for low-selectivity queries and an HNSW-based filtered ANNS algorithm for other selectivity levels.

    \item 
    We devise an inline-filtering search algorithm integrating HNSW with a \textit{exclusion distance } mechanism. 
    This mechanism adaptively repels filtering-violated vectors from the query while promoting valid candidates along the search path, thereby improving efficiency without sacrificing recall.

    \item 
    Our selectivity-driven search selector estimates query selectivity on the fly, dynamically routing each query to the optimal algorithm. This ensures stable performance under arbitrary filtering conditions by choosing between the pre-filtering brute-force method and our proposed inline-filtering search algorithm.

    \item 
    We conduct extensive experiments on real-world datasets with various filtering conditions. \ourmethod\ achieves a 1.3–5$\times$ improvement in QPS at $Recall@10 = 95\%$ compared to filter-agnostic approaches while exhibiting comparable performance compared to state-of-the-art tailored approaches in some filtering conditions.
\end{itemize}

\zl{
We integrate FAVOR into HAKES~\cite{hakes}\footnote{\url{https://github.com/nusdbsystem/HAKES}}, a scalable and modular vector database for supporting AI workloads.
FAVOR enables HAKES to efficiently support hybrid vector–attribute queries with arbitrary filtering conditions, providing stable performance across varying query selectivities.
}

The paper is structured as follows. Section~\ref{sec:pre} provides the background on Filtered ANNS and state-of-the-art methods. Section~\ref{sec:overview} discusses the design guidelines and provides an overview of \ourmethod. Section~\ref{sec:policy} and Section~\ref{sec:saed} detail the core components, the selectivity-driven search selector, and the \ouralg search algorithm. Section~\ref{sec:evaluation} evaluates \ourmethod. Section~\ref{sec:related-work} reviews related works and Section~\ref{sec:conclusion} concludes.

%% file: Preliminary.tex
\section{PRELIMINARY AND BACKGROUND}
\label{sec:pre}

We first establish the notation and formally define the problem of filtered ANNS. Table~\ref{tab:notation} lists the key symbols used in this paper.

\begin{table}[t]
\centering
\caption{Key notations.}
\vspace{-2mm}
\label{tab:notation}
\scalebox{0.98}{
\begin{tabular}{l|l}
\hline
\textbf{Symbol} & \textbf{Description} \\
\hline
$\mathcal{\hat{D}}$ & Vector set\\
$A$ & Attribute set \\
$\mathcal{D}$ & Vector dataset with $A$ \\
$\mathcal{F}$ & Set of filtering conditions\\
$\bm{q}$ & Query vector in $\mathbb{R}^l$ \\
$\bm{v}^{T}$ / $\bm{v}^{N}$ & Vector whose $A \in \mathcal{F}$ / $A \notin \mathcal{F}$\\
TD / NTD & Set of ($\bm{v}_{i}^{T}$, $A_i$) / ($\bm{v}_{j}^{N}$, $A_j$) in $\mathcal{D}$ for a ($\bm{q}$, $\mathcal{F}$)\\
$\mathcal{V}$ & Visited list: set of vectors already visited\\
$\mathcal{C}$ & Candidate set: set of vectors to be explored\\
$\mathcal{R}$ & Result set: the $\mathit{ef}$ nearest vectors to $\bm{q}$ in $\mathcal{C}$\\
$\mathcal{S}$ & Target result set: the $k$ nearest vectors to $\bm{q}$ in $\mathcal{R}$\\
$p$ & Selectivity: proportion of TD in $\mathcal{D}$\\
\hline
\end{tabular}
}
\end{table}

\subsection{Definition}
\label{sec:2-fa}
\subsubsection{Approximate Nearest Neighbor Search (ANNS)} For a query vector $\bm{q}$ and a vector dataset $\mathcal{\hat{D}}=\left\{\bm{v_{1}}, \bm{v_{2}}, \ldots, \bm{v_{\mathrm{n}}}\right\}\in \mathbb{R}^l$, where $l$ is the dimensionality of the vector space, the purpose of approximate nearest neighbor search is to find the $k$ approximate nearest vectors to $\bm{q}$. ANNS addresses this task by using index structures (\eg, hash\cite{LSH1,LSH2,LSH3}, tree\cite{tree1,tree2,tree3,tree4}, graph\cite{HNSW,NSW}) to achieve sublinear search time while trading off minimal accuracy for efficiency gains. 

\subsubsection{Filtered ANNS}
Let each vector $\bm{{v}_i}\in \mathcal{\hat{D}} (i = 1, 2, \ldots, n)$ be associated with a ${m}$-dimensional scalar attribute set ${A}_i = \left\{a_{i 1}, a_{i 2}, \ldots, a_{i m}\right\}$, where each ${a}_{ij} (j = 1, 2, \ldots, m)$ is a scalar value. 
The vector dataset with attributes can be formally defined as \\ $\mathcal{D}=\left\{\left(\bm{v_{1}}, A_{1}\right),\left(\bm{v_{2}}, A_{2}\right), \ldots,\left(\bm{v_{n}}, A_{n}\right)\right\}$.
Given a query vector $\bm{q}$ and a set of filtering conditions $\mathcal{F}$, the target subset is defined as
\begin{equation*}
\mathcal{S}=\left\{\left(\bm{v_{i}}, A_{i}\right) \in \mathcal{D} \mid A_{i} \in \mathcal{F}\right\}.
\end{equation*}

Filtered ANNS aims to acquire the approximate $k$ nearest vectors to $\bm{q}$ in $\mathcal{S}$. This paper focuses on filtered ANNS with arbitrary $\mathcal{F}$. 
Generally, the $\mathcal{F}$ can be categorized into four primary types, including \textit{Equality}~\cite{HQANN,NHQ,CAPS,ung,Filtered-DiskANN}, \textit{Inclusion}~\cite{ung,Filtered-DiskANN}, \textit{Range}~\cite{SeRF,window,iRangeGraph,UNIFY,ARKGraph}, and \textit{Logic}~\cite{AIRSHIP,ACORN,SIEVE,NaviX}.

\subsection{Hierarchical Navigable Small World (HNSW)}
HNSW~\cite{HNSW} is an efficient graph-based indexing approach.
It constructs a multi-layered proximity graph that supports logarithmic-scale search complexity.
The index comprises $L+1$ layers (from layer $L$ down to layer $0$), where higher layers contain sparsely connected nodes functioning as ``highways'' across the data space, while the base layer (layer $0$) maintains detailed local neighborhood structures for precise retrieval. 

During the index construction phase, each vector is inserted into the base layer and added to higher layers with exponentially decreasing probability. At each layer, it connects to its approximate nearest neighbors. 
Specifically, for a vector to insert $\bm{v}$, the topmost layer level $l$ is sampled from an exponentially decaying probability distribution. 
Then, starting from an entry point designated at the top layer, greedy search is performed to locate the nearest vectors to $\bm{q}$, and the nearest $\bm{v}$ serves as the entry point for the next layer. Iteratively, the neighbors at layer $l$ to $L$ are identified and connections are built with $\bm{v}$ being inserted to those layers.

During the query search phase, the search starts from a designated entry point at the top layer and navigates downward through successive layers, progressively refining a candidate set, $\mathcal{C}$, until reaching the base layer, where the final $k$ nearest neighbors are identified.
Specifically, the algorithm executes $\textit{GreedySearch}(\bm{q}, 1, ep, \mathcal{G})$ at each non-base layer to iteratively select the optimal entry node $ep$ for transitioning to the subsequent layer.
Once reaching the base layer, the algorithm transitions from single-node selection to $\textit{GreedySearch}(\bm{q}, \mathit{ef}, ep, \mathcal{G})$, where $\mathit{ef} > k$ is to ensure comprehensive exploration of the vector space.
Finally, the top-$k$ vectors nearest to the $\bm{q}$ in $\mathcal{R}$ are selected as the query result.

\begin{algorithm}[t]
\caption{$\text{GreedySearch}(\bm{q},\mathit{ef},ep,\mathcal{G})$}\label{alg:greedysearch}
\KwIn{$\text{query vector } \bm{q}, \text{entry point } ep, \text{graph index } \mathcal{G}$}
\KwOut{$\text{result set of } \mathit{ef} \text{ approximate nearest neighbors }\mathcal{R}$}
$\mathcal{V} \gets ep$ \tcp*{visited list}
$\mathcal{C} \gets ep$ \tcp*{candidate set}
$\mathcal{R} \gets ep$ \tcp*{result set}

\While{$\mathcal{C} \neq \emptyset$}{
    $\bm{v}_{\alpha} \gets 
    \arg \min _{\bm{v} \in \mathcal{C}} Dis(\bm{q}, \bm{v})$\;
    $\bm{v}_{\beta} \gets \arg \max _{\bm{v} \in \mathcal{R}} Dis(\bm{q}, \bm{v})$\;
    \If{$Dis(\bm{q},\bm{v}_{\alpha})>Dis(\bm{q},\bm{v}_{\beta})$}{
        \textbf{break}\;
    }
    \ForEach{$\bm{v} \in \text{neighbors of } \bm{v}_{\alpha}$}{
        \If{$\bm{v} \notin \mathcal{V}$}{
            $\mathcal{V} \leftarrow \mathcal{V} \bigcup \bm{v}$\;
            $\bm{v}_{\beta} \gets \arg \max _{\bm{v} \in \mathcal{R}} d(\bm{q}, \bm{v})$\;
            \If{$(Dis(\bm{q}, \bm{v})<Dis(\bm{q},\bm{v}_{\beta}))\parallel(\left | \mathcal{R} \right | <\mathit{ef})$}{
            $\mathcal{C} \leftarrow \mathcal{C} \bigcup \bm{v}$\;
            $\mathcal{R} \leftarrow \mathcal{R} \bigcup \bm{v}$\;
                \If{$\left | \mathcal{R} \right | > \mathit{ef}$}{
                    $\mathcal{R} \gets \mathcal{R} \setminus \arg \max _{\bm{v} \in \mathcal{R}} Dis(\bm{q}, \bm{v})$\;
                }
            }
        }
    }
}
\Return $\mathcal{R}$\;
\end{algorithm}

\begin{table}[h]
\caption{Summary table w.r.t filtered ANNS solutions.}
\vspace{-2mm}
\label{tab:comparison}
\resizebox{\textwidth}{!}{
\begin{tabular}{c|c|c|c|c|c|c|c|c|c}
    \toprule 
    \multirow{2}{*}{\textbf{Methods}} & \multirow{2}{*}{\textbf{Graph Type}} & \multicolumn{4}{c|}{\textbf{Filter Support}} & \multirow{2}{*}{\makecell{\bf Vectors on \\ \bf search path}} & \multirow{2}{*}{\makecell{\bf Termination \\ \bf efficiency }} & \multirow{2}{*}{\textbf{General QPS}} & \multirow{2}{*}{\makecell{\bf QPS under \\ \bf low selectivity}} \\
     &   & \textbf{Equality}   & \textbf{Inclusion}   & \textbf{Range}   & \textbf{Logic}  &  &  &  &     \\ 
    \midrule
    \bf SeRF\cite{SeRF} &  Attribute-tailored  &  \ding{53} & \ding{53}  &  \ding{51} & \ding{53} & TD  & High  &  High  &  High  \\
    \hline
    
    \bf UNG\cite{ung} &  Attribute-tailored  &  \ding{51} & \ding{51}  &  \ding{53} & \ding{53} & TD  & High  &  High  &  High  \\

    \hline

    \bf ACORN-$\gamma$\cite{ACORN} &  Dense-Conventional  &  \ding{51} & \ding{51}  &  \ding{51} & \ding{51} & TD  & Medium  &  Medium  &  Low  \\
    \hline
    
    \bf RSF\cite{HNSW} &  Conventional  &  \ding{51} & \ding{51}  &  \ding{51} & \ding{51} & TD \& NTD  & Low  &  Medium  &  Low  \\
    \hline
    \bf \ourmethod  &  Conventional  &  \ding{51} & \ding{51}  &  \ding{51} & \ding{51} & TD \& NTD  & High  &  High  &  High  \\
\bottomrule
\end{tabular}
}
\end{table}

In addition to the concepts presented in Table~\ref{tab:notation}, we focus on two key concepts of HNSW that are crucial for our analysis and the improvement of existing limitations.
\begin{itemize}[noitemsep,topsep=0pt,parsep=0pt,partopsep=0pt,leftmargin=*]
\item \textbf{Search Path}: the sequence of data selected to extend the traversal path during the search process, starting from an entry point, along with the data points whose vectors are closer to the $\bm{q}$ compared to their neighbor vectors. 
\item \textbf{Termination Condition}: the criterion for the termination of the search process. In HNSW, termination is triggered when all vectors in $\mathcal{C}$ are farther from $\bm{q}$ than all vectors in $\mathcal{R}$, or when $\mathcal{C}$ is empty.
\end{itemize}

\subsection{Variants for filtered ANNS}
\label{sec:var}
Typically, the search process of HNSW could be extended with an inline-filtering approach to support filtered ANNS without modifying the graph construction process.
Meanwhile, there also exist solutions that integrate the attributes with the proximity graph to achieve high efficiency.

\subsubsection{Representative Approaches}
We summarize four representative variants of HNSW-based inline-filtering approaches that inspired our work.
Approaches such as \textit{UNG}~\cite{ung} and \textit{SeRF}~\cite{SeRF} support filtering for specialized condition sets $\mathcal{F}$, while \textit{ACORN}~\cite{ACORN} and \textit{Result-Set-Filtering}~\cite{HNSW} enable filtering for arbitrary $\mathcal{F}$.
All of them adopt termination conditions similar to those in HNSW.

\begin{itemize}[noitemsep,topsep=0pt,parsep=0pt,partopsep=0pt,leftmargin=*]
\item \textbf{UNG} constructs a directed acyclic Label Navigation Graph (LNG) to organize label semantics.
It builds a local proximity graph for each label set and interconnects them into a global index through cross-group edges derived from LNG containment relationships.
At query time, UNG locates the minimal superset covering the query label set within the LNG and initiates graph traversal from the corresponding vector nodes.
Guided by cross-group edges, the traversal is restricted to the subgraph that satisfies the filter condition, ensuring that $\mathcal{C}$ and $\mathcal{R}$ consist solely of TD.

\item \textbf{SeRF} constructs an HNSW graph by inserting nodes in ascending order of attribute values and assigning validity intervals to edges using insertion or deletion timestamps.
This design compresses multiple range-specific indexes into a single unified structure.
For a given query, SeRF dynamically prunes edges whose validity intervals fall outside the query’s range boundaries, forming a tailored subgraph for efficient filtered traversal.
Consequently, its $\mathcal{C}$ and $\mathcal{R}$ all contain only TD.

\item \textbf{ACORN-$\gamma$} performs filtered ANNS by augmenting a dense HNSW index.
During index construction, it introduces additional neighbors for each data point to enhance connectivity.
At query time, the algorithm filters out NTD neighbors but computes distances only between $\bm{v}^{T}$ and $\bm{q}$.
The TD point with the shortest distance extends the search path, effectively forming a subgraph composed entirely of TD.

\item \textbf{RSF} is identical to HNSW in index construction and termination conditions. The only difference is that it only permits TD to enter $\mathcal{R}$ during the query search process. 
\end{itemize}

\subsubsection{Discussion and Rethinking}
Table~\ref{tab:comparison} summarizes the performance ratings and search strategies of representative methods.
Several consistent observations can be drawn.
By comparing the columns ``Graph Type'' and ``Filter Support'', it becomes evident that attribute-tailored graphs generally achieve better performance but are limited to specific types of filtering conditions.
This limitation arises because the topological structures of these tailored indexes vary across filter types, preventing them from generalizing to other filtering conditions.
Consequently, UNG cannot support range filtering, while SeRF is only applicable to range-based filters.
In contrast, ACORN-$\gamma$ and RSF, which employ conventional graph structures, are capable of handling arbitrary filtering conditions.
However, this generality comes at the cost of lower efficiency compared to tailored methods.

It can also be seen from the table that most approaches adopt a TD-preferred strategy, where distance computations are performed only between $\bm{v}^{T}$ and $\bm{q}$, and only TD are collected into $\mathcal{C}$.
While this design reduces redundant similarity evaluations, it also requires a tailored index structure to maintain sufficient graph connectivity.
Without such structural support, connectivity becomes fragile, leading to inefficiency in search.
For instance, ACORN-1, an extreme setting for ACORN-$\gamma$, often encounters cases where no TD exist within the neighborhood or second-order neighborhood of the current node.
Although this issue could be alleviated by increasing node out-degrees in ACORN-$\gamma$, it introduces additional overhead in both index construction and neighbor exploration.
In contrast, RSF avoids connectivity loss by always extending the search path toward the nearest neighbor to $\bm{q}$, regardless of whether the neighbor is TD or NTD.
However, since $\mathcal{C}$ is frequently populated with NTD close to $\bm{q}$ while $\mathcal{R}$ accepts only TD, RSF struggles to terminate efficiently.
It must locate TD sufficiently close to $\bm{q}$, leading to unnecessary distance computations for NTD.
Thus, the tradeoff between the graph connectivity and efficiency is a crucial design point for HNSW-based solutions to filtered ANNS.

The table also reveals an interesting phenomenon: the absence of attribute-tailored index structures leads to unstable performance when selectivity is low. 
In practical scenarios, queries are continuously issued with diverse filtering conditions, leading to highly variable selectivity levels.
When a low-selectivity query triggers throughput degradation, it can propagate delays throughout the request queue, causing a sharp rise in end-to-end latency and severely impairing user experience.
This highlights the necessity of addressing low-selectivity robustness as a fundamental requirement for real-world ANNS systems.

In summary, while each search strategy demonstrates strengths under certain conditions, ensuring stable and efficient performance across dynamic query selectivity and variant filter types remains an open research problem that motivates our work.

%% file: FAVOR.tex
\section{\ourmethod}
\label{sec:overview}
We present the design guidelines and the overview of our work in this section.

\subsection{Key idea} 
\label{sec:motivation}

\begin{figure*}[t]
    \centering
    \includegraphics[width=0.98\linewidth]{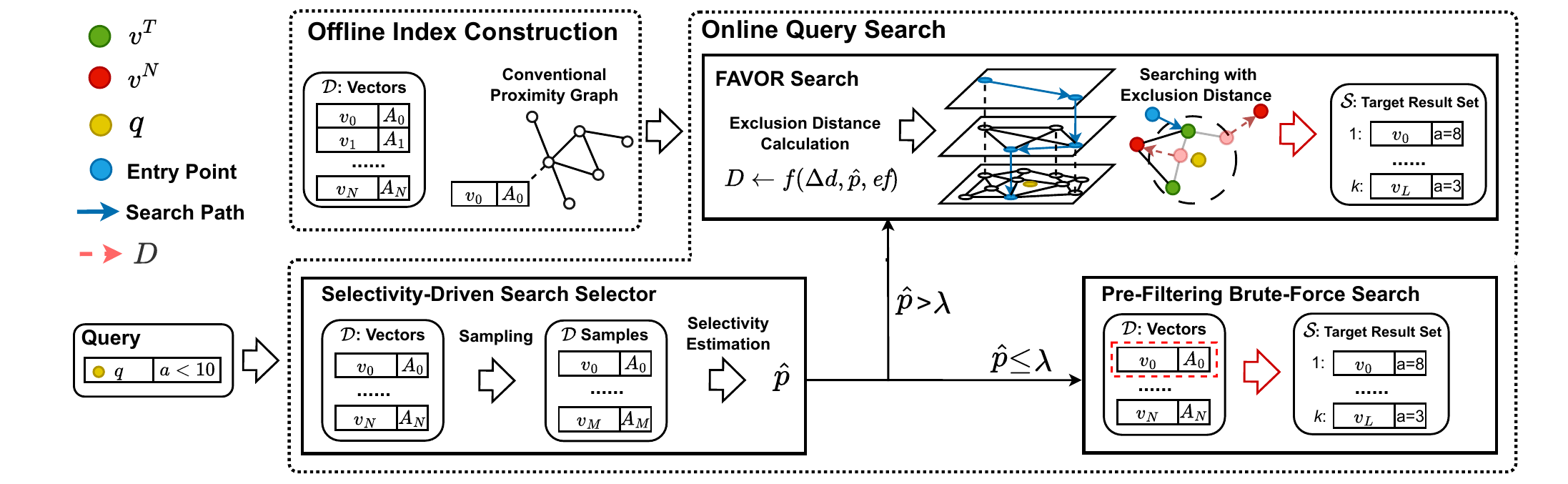}
    \vspace{-4mm}
    \caption{Workflow of \ourmethod.}
    \vspace{-2mm}
    \label{fig:overview}
\end{figure*}

We aim to develop an efficient filtered ANNS method that supports arbitrary $\mathcal{F}$ while maintaining high QPS across the variant selectivity range.

\subsubsection{Key Idea.} \label{sec:keyidea}
Our key idea is that high performance for filtered ANNS within the HNSW framework can be achieved by increasing the number of TD points that are closer to $\bm{q}$ than their neighbors on the search path.
Existing approaches either construct tailored graph structures (\eg, UNG~\cite{ung} and SeRF~\cite{SeRF}) or modify search strategies (\eg, ACORN~\cite{ACORN}) to favor TD points. 
Tailored structures explicitly connect TD points and ensure traversal only among them, thereby maximizing proximity to $\bm{q}$ but sacrificing flexibility across different filtering conditions.
Conversely, passive search modification strategies, such as those used in ACORN, force the selection of TD points to extend the search path and enlarge node out-degrees to increase TD encounter probability. 
However, they cannot ensure that TD points are actually closer to $\bm{q}$ than their neighbors, leading to insufficient exploration and degraded accuracy.

Our approach aims to balance two essential objectives: actively prioritizing TD points during search and comprehensively exploring the query neighborhood, while preserving full flexibility for arbitrary filtering conditions. 
The key lies in dynamically reshaping the distance distribution during search to make TD points closer to $\bm{q}$ than their neighbors, thereby achieving high performance without sacrificing generality.

\subsubsection{Design Guidelines.}
Building upon the discussion and rethinking in Section~\ref{sec:var}, we summarize several key design guidelines to guide the development of our method:
\begin{itemize}[noitemsep,topsep=0pt,parsep=0pt,partopsep=0pt,leftmargin=*]
\item \textbf{G.1:} Employ a general proximity graph to organize data and perform query search, ensuring adaptability to arbitrary filtering conditions.
\item \textbf{G.2:} Compute distances between all neighbors and $\bm{q}$ so that the search path always extends through the data point nearest to $\bm{q}$.
\item \textbf{G.3:} Allow NTD points to enter $\mathcal{R}$ to enable accurate and efficient termination decisions.
\item \textbf{G.4:} Identify the selectivity of each query and adapt the search strategy to improve efficiency in low-selectivity scenarios.
\end{itemize}
\subsection{Design of \ourmethod{}}
We propose \ourmethod{} based on the guidelines \zl{and our key idea}. 
It estimates selectivity and selects a suitable search strategy correspondingly (\textbf{G.4}). 
Based on the standard HNSW graph and query search strategy, it requires no knowledge of queries beforehand and supports arbitrary $\mathcal{F}$ at predetermined recall rates (\textbf{G.1-G.3}). 
Note that it enhances the search algorithm by dynamically making TD closer to the $\bm{q}$ compared to their neighbors.

\subsubsection{Architecture} 
As shown in Figure~\ref{fig:overview}, \ourmethod\ follows the two-phase search paradigm as vanilla HNSW: offline index construction and online query search.
The offline phase prepares a general proximity graph without any filtering-specific design, while the online phase adaptively executes one of two search algorithms based on the estimated query selectivity. 
The core of \ourmethod\ lies in its selectivity-driven search strategy and the two complementary algorithms that ensure stable, high-performance ANNS across diverse filtering conditions.
\begin{itemize}[noitemsep,topsep=0pt,parsep=0pt,partopsep=0pt,leftmargin=*]

\item \textit{Selectivity-driven search selector.} The selector first conducts a sampling on the vector dataset $D$ and then estimates the query selectivity. 
It determines the appropriate search mode to balance efficiency and accuracy. 
The estimation and selection strategy details are described in Section~\ref{sec:policy}.

\item \textit{\ouralg search algorithm.} The algorithm is an HNSW-based inline-filtering ANNS method with dynamic distance computation.
\textcolor{black}{
An exclusion distance mechanism achieves this dynamic computation during the search process.
}
\ouralg search algorithm is activated to perform a focused and efficient vector search over the most relevant index regions when the query exhibits qualified selectivity.
The details of the search process are presented in Section~\ref{sec:saed}.

\item \textit{Pre-filtering brute-force search algorithm.}
The algorithm is a linear scanning search method.
That is, a pre-filtering step narrows the candidate space before applying a lightweight brute-force search, ensuring comprehensive yet efficient retrieval for low-selectivity queries.
The algorithm will be introduced in Section~\ref{sec:policy}.

\end{itemize}

\subsubsection{Processing Workflow} 
We explain the processing workflow of \ourmethod in its two phases below.

In the offline index construction phase, \ourmethod\ builds an index over the vectors in $\mathcal{D}$ using a conventional graph connectivity algorithm~\cite{RNG}.
Each vector is stored together with its attribute set $A$ to support efficient inline filtering at query time.
During construction, \ourmethod\ records the average pairwise distance between vectors, denoted as $\Delta d$ (see Section~\ref{sec:ed}), which captures the intrinsic distance distribution of $\mathcal{D}$.
This statistic is stored in the index metadata and later used to compute exclusion distances and guide search path optimization during online query processing.
Note that this phase is performed entirely offline and does not rely on any prior query information.

In the online query search phase, the incoming query $\bm{q}$ and its filtering condition $\mathcal{F}$ are first processed by the selectivity-driven search selector.
This component estimates the query selectivity $\hat{p}$ based on $\mathcal{F}$ (see Section~\ref{sec:estimation}) and routes the query according to a predefined threshold $\lambda$ (see Section~\ref{sec:sthreshold}).
If $\hat{p} < \lambda$, the system executes Pre-Filtering Brute-Force Search, which performs a lightweight scan over a reduced candidate set to return the top-$k$ results.
Otherwise, the \ouralg search algorithm is activated.
\ourmethod\ computes the exclusion distance using the pre-recorded $\Delta d$ and the estimated $\hat{p}$ (see Section~\ref{sec:ed}).
It then performs a guided traversal over the HNSW index based on this exclusion distance until the termination criteria are met (see Section~\ref{sec:tc}), after which the top-$k$ results are returned.

%% file: Selector.tex
\section{SELECTIVITY-DRIVEN SEARCH SELECTOR}
\label{sec:policy}

\subsection{Selector Design by Selectivity}
\label{sec:sthreshold}
HNSW-based methods are highly sensitive to query selectivity $p$.
When $p$ is high, a dense distribution of TD allows the navigation-based search mechanism to efficiently locate the top-$k$ data points nearest to the $\bm{q}$.
However, when $p$ is low, TD becomes sparsely distributed, causing the search to diverge into irrelevant regions of the graph.
This leads to numerous ineffective visits to NTD and severely reduced search efficiency.
In contrast, 
\textcolor{black}{the pre-filtering brute-force approach (PreFBF), which applies filtering before similarity computation, could perform more effectively under low-selectivity conditions~\cite{SeRF}.
}
 
To adaptively select the two search strategies, we introduce a selectivity threshold $\lambda$ that determines when to use PreFBF or the HNSW-based algorithm (see Section \ref{sec:saed}).
A conservative threshold ensures that PreFBF is chosen when selectivity falls below $\lambda$, while the HNSW-based algorithm is used otherwise. 
We determine the $\lambda=1\%$ based on empirical measurement.
We refer to the interval in which $p$ is close to but slightly exceeds 
1\%
as the middle selectivity range. 
Since it is a less representative scenario~\cite{ACORN,ung}, we adopt a conservative retrieval strategy that prioritizes graph-based search methods because they incur lower potential error costs and thus provide more robust retrieval quality (see Section~\ref{subsec:perf}).

\subsection{Selectivity Estimation}
\label{sec:estimation}
When a query $q$ arrives, we estimate $p$ first to determine the corresponding search strategy.
A straightforward way to obtain accurate selectivity is to scan the entire dataset and count TD points. 
Yet, it is costly and cannot scale with the dataset size.
Given that the estimation is an online operation, \ourmethod prioritizes efficiency and generality over accuracy and employs a sampling approach. 
Given a query, we collect the proportion of TD points in a randomly drawn subset from the dataset and express it by $\hat{p}$. 
To assess the reliability of the estimation method, we discuss the relative error of the random sampling approach.
Since the number of TD points follows a hyper-geometric distribution~\cite{feller1991introduction}, the relative error of $\hat{p}$ is
\begin{equation}
\label{eq:error}
    \textit{Relative Error} = \frac{\sqrt{\text{Var}(\hat{p})}}{p} = \sqrt{ \frac{(1-p)}{n p} \left(1 - \frac{n}{N}\right) },
\end{equation}
where $n$ is the sample size and $N$ is the size of the total dataset. 
The relative error decreases as the sample size $n$ and selectivity $p$ increase. 
For million-scale datasets, the relative error remains around 1\% even when $p\approx 1\%$ with a sampling rate of $n/N = 1\%$. 
The error only becomes significant when $p<<1\%$, for which the strategy selects PreFBF as the search strategy based on $\lambda=1\%$. 
Since PreFBF's search process does not require $\hat{p}$, it does not affect search performance. 

%% file: FAVORsearch.tex
\begin{figure}[t]
    \centering
	\includegraphics[width=0.62\linewidth]{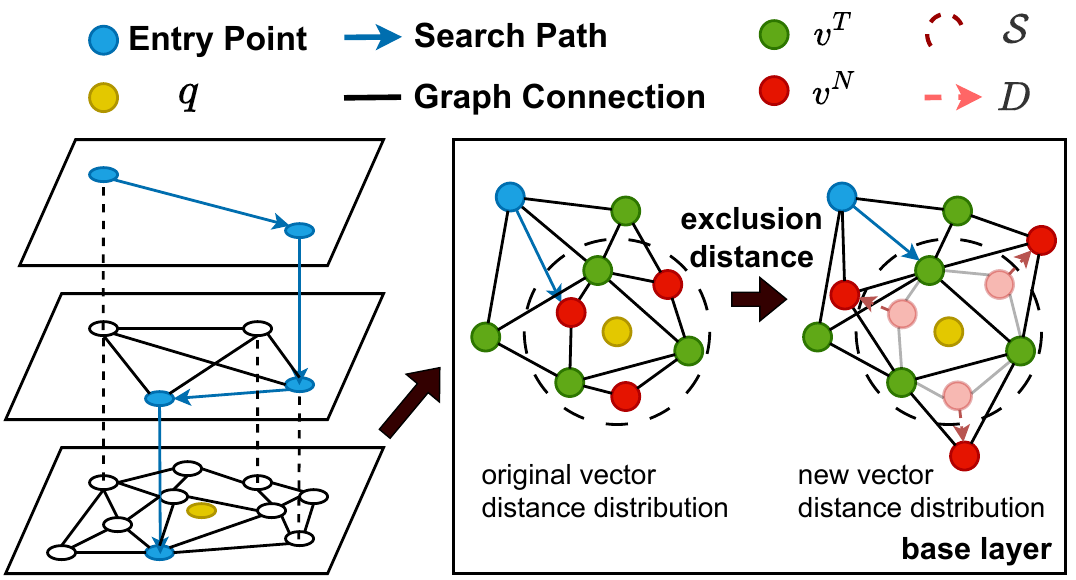}
    \caption{\ouralg search process.}
    \label{fig:SAED}
\end{figure}

\section{\ouralg SEARCH ALGORITHM}
\label{sec:saed}
In this section, we detail the basic idea of \ouralg search algorithm, the filtered search workflow, the determination of exclusion distance, and the optimization of termination conditions.

\subsection{Basic Idea}

Existing HNSW-based filter-agnostic ANNS methods treat the inter-vector distance distribution as static during search, preventing them from achieving the essential goal of making TD appear closer to the query vector $\bm{q}$ than their NTD neighbors.
While tailored approaches explicitly reshape this distribution through index reconstruction during the pre-search process, such modifications are infeasible for filter-agnostic designs.
Therefore, our idea is to dynamically adjust the distribution during the retrieval process itself to support different filter conditions.
The key intuition is to perform real-time distance correction for each visited vector based on its compliance with the filtering condition, thereby continuously reshaping the target distance distribution to emulate the behavior of tailored indexes.
As the central innovation of this work, we introduce an exclusion distance mechanism that penalizes non-target data (NTD), enabling this adaptive correction in an online manner.

\subsubsection{Definition.} In our approach, the distance function $Dis(\bm{q}, \bm{v})$ represents the Euclidean distance between two vectors. For filtered ANNS, we first utilize $\mathcal{F}$ to distinguish TD and NTD. 
To reduce the impact of NTD during search, we impose an exclusion distance that pushes NTD farther from $\bm{q}$, thereby increasing the likelihood that nearby TD points become the top candidates.
Formally, the distance function for the exclusion distance mechanism is formulated as
\begin{equation}
\label{eq:inidistance}
\overline{Dis}\left(\bm{q}, \bm{v}\right)=\left\{\begin{array}{ll}
Dis(\bm{q}, \bm{v}^{T}), & A \in \mathcal{F}, \\
Dis(\bm{q}, \bm{v}^{N}) + D, & A \notin \mathcal{F},
\end{array}\right.
\end{equation}
where $\overline{Dis}(\cdot)$ denotes the adjusted distance after distribution correction, $D$ is the exclusion distance, $\bm{v}$ represents a visited data point, and $A$ denotes its associated attribute set.

\subsubsection{Illustration.}
As illustrated in the base layer of Figure~\ref{fig:SAED}, under the original vector distance distribution, a NTD data point ($\bm{v}^{N}$), colored in red, is likely to be selected to extend the search path.
After applying the exclusion distance, however, the NTD points are effectively pushed farther from the query $\bm{q}$, making a TD point ($\bm{v}^{T}$) more likely to be selected instead.
This adjustment brings TD points closer to $\bm{q}$ than their neighboring NTD, enabling more efficient search expansion and potentially yielding higher query throughput (QPS).
Building on this intuition, we develop \ouralg search algorithm, which operates on the standard HNSW graph without modifying its construction process. 
Instead, our algorithm dynamically redefines the distances between $\bm{q}$ and visited data points during search and refines the termination condition, thereby significantly enhancing search efficiency.

\subsection{Search Workflow}
\label{sec:saed-wf}

Based on the offline-built index, the overall search workflow is depicted in Figure~\ref{fig:SAED} and Algorithm~\ref{alg:saed}. 
The process starts from the entry point at the top layer and proceeds layer by layer, where a greedy search algorithm identifies the data point nearest to the query $\bm{q}$ within the current layer.
This point then serves as the entry point for the next lower layer.
During this hierarchical descent, attribute filtering is temporarily omitted to enable rapid convergence toward the region near $\bm{q}$.
Once the entry point reaches the bottom layer, the \textit{OptiGreedySearch} procedure (Algorithm~\ref{alg:optigreedy}) is invoked to populate $\mathcal{C}$ and $\mathcal{R}$.
Finally, the top-$k$ TD points nearest to $\bm{q}$ are selected from $\mathcal{R}$ to construct the $\mathcal{S}$.

\begin{algorithm}[t]
\caption{$\text{FAVORSearch}(q,p,k,\mathcal{G})$}\label{alg:saed}
\KwIn{$\text{query vector } \bm{q}, \text{selectivity }p, \text{graph index } \mathcal{G}$}
\KwOut{$\text{result set of } k \text{ approximate nearest neighbors }$}
$\mathcal{R} \gets \emptyset$ \tcp*{result set}
$\mathcal{S} \gets \emptyset$ \tcp*{target result set}

$ep \gets \text{entry point in } \mathcal{G}$\;
$L \gets \text{number of layers in } \mathcal{G}$\;
\For{$l=L-1$ \KwTo $1$}{
    $\mathcal{R} \gets \text{GreedySearch}(\bm{q},1,ep,\mathcal{G}_l)$\;
    $\mathit{ep} \gets \arg \min _{\bm{v} \in \mathcal{R}} Dis(\bm{q}, \bm{v})$\;
}
$\mathcal{R} \gets \text{OptiGreedySearch}(\bm{q},\mathcal{F},\mathit{ef},\mathit{ep},\mathcal{G}_0)$\;
$\mathcal{S} \gets k \text{ nearest vectors to } \bm{q} \text{ in } \mathcal{R}$\;

\Return $\mathcal{S}$\;
\end{algorithm}

Algorithm~\ref{alg:optigreedy} describes the process of how \ouralg implements search using the exclusion distance at the base layer. The algorithm employs an array, a min-heap, and a max-heap to maintain $\mathcal{V}$, $\mathcal{C}$, and $\mathcal{R}$, respectively, while initializing a count of TD in $\mathcal{R}$. Note that the order of vectors in $\mathcal{C}$ and $\mathcal{R}$ is maintained according to the vector distance distribution with an added exclusion distance. Consequently, when a data point enters them, it is first checked against $\mathcal{F}$, then the distance is calculated based on Equation (\ref{eq:inidistance}). 

In the beginning, the entry point enters into $\mathcal{V}$, $\mathcal{C}$ and $\mathcal{R}$. When $\mathcal{C}$ is not empty, $\mathcal{C}$'s root node, \ie, the nearest data point to $\bm{q}$ in $\mathcal{C}$ under the new vector distance distribution, is removed from $\mathcal{C}$ and used to extend the search path. If $\mathcal{C}$'s root node is farther from $\bm{q}$ compared to $\mathcal{C}$'s root node, \ie, the farthest data point to $\bm{q}$ in $\mathcal{R}$ under the new vector distribution, while the count of TD in $\mathcal{R}$ is more than $\mathit{ef}$/2, the search process terminates. The top-$k$ nearest vectors to $\bm{q}$ in $\mathcal{R}$ are the results. Algorithm~\ref{alg:optigreedy} (line 1-9) outlines the above process.

When the termination condition is not satisfied, the algorithm begins to traverse the neighbors of the point used to extend the path. If a neighbor hasn't been previously visited, it is added to $\mathcal{V}$. Then, it is added to $\mathcal{C}$ and $\mathcal{R}$ if it is closer to $\bm{q}$ than the current root of $\mathcal{R}$, or $\mathcal{R}$ is not full. 
In this case, if its attributes meet $\mathcal{F}$, the count of TD in $\mathcal{R}$ adds 1. 
Furthermore, when the population operation makes the size of $\mathcal{R}$ beyond $\mathit{ef}$, the root node of $\mathcal{R}$ is removed from $\mathcal{R}$. If this removed data point's attributes meet $\mathcal{F}$, the count of TD in $\mathcal{R}$ subtracts 1. The above process iterates until all neighbors have been explored. Algorithm~\ref{alg:optigreedy} (line 10-24) outlines the above process. During this process, we can enable more TD points to pave the search path.  

\begin{algorithm}[t]
\caption{$\text{OptiGreedySearch}(\bm{q},\mathcal{F},\mathit{ef},\mathit{ep},\mathcal{G})$}\label{alg:optigreedy}
\KwIn{$\text{query vector } \bm{q}, \text{filtering conditions }\mathcal{F}, \text{entry point } \mathit{ep},$ \\ 
$\text{graph index } \mathcal{G}$}
\KwOut{$\text{result set of } \mathit{ef} \text{ approximate nearest neighbors }$}
$\mathcal{V} \gets \mathit{ep}$ \tcp*{visited list}
$\mathcal{C} \gets \mathit{ep}$ \tcp*{candidate set}
$\mathcal{R} \gets \mathit{ep}$ \tcp*{result set}
$n \gets 0$\ \tcp*{count for TD in $\mathcal{R}$}

\While{$\mathcal{C} \neq \emptyset$}{
    $\bm{v}_{\alpha} \gets \text{extract } \arg\min _{\bm{v} \in \mathcal{C}} \overline{Dis}(\bm{q},\bm{v})$\;
    $\bm{v}_{\beta} \gets \arg\max _{\bm{v} \in \mathcal{R}} \overline{Dis}(\bm{q},\bm{v})$\;
    
    \If{$\overline{Dis}(\bm{q},\bm{v}_{\alpha})>\gamma \cdot \overline{Dis}(\bm{q},\bm{v}_{\beta})$ and $\overline{p}>0.5$}{
        \textbf{break}\;
    }
    \ForEach{$\bm{v} \in \text{neighbors of }\bm{v}_{\alpha}$}{
        \If{$\bm{v} \notin \mathcal{V}$}{
            $\mathcal{V} \leftarrow \mathcal{V} \bigcup \bm{v}$\;
            $\bm{v}_{\beta} \gets \arg\max _{\bm{v} \in \mathcal{R}} \overline{Dis}(\bm{q}, \bm{v})$\;
            $A \gets \text{attribute of } \bm{v}$\;
            \If{$(\overline{Dis}(\bm{q}, \bm{v})<\overline{Dis}(\bm{q},\bm{v}_{\beta}))\parallel(\left | \mathcal{R} \right | <\mathit{ef})$}{
                $\mathcal{C} \leftarrow \mathcal{C} \bigcup \bm{v}$\;
                $\mathcal{R} \leftarrow \mathcal{R} \bigcup \bm{v}$\;
                \If{$A \in \mathcal{F}$}{
                    $n \gets n+1$\;
                }
                \If{$\left | \mathcal{R} \right | > \mathit{ef}$}{
                    $A \gets \text{attribute of } \arg\max _{\bm{v} \in \mathcal{R}} \overline{Dis}(\bm{q},\bm{v})$\;
                    $\mathcal{R} \gets \mathcal{R} \setminus \arg\max _{\bm{v} \in \mathcal{R}} \overline{Dis}(\bm{q}, \bm{v})$\;
                    \If{$A \in \mathcal{F}$}{
                        $n \gets n -1$\;
                    }
                }
            }
        }
    }
}
\Return $\mathcal{R}$\;
\end{algorithm}

\subsection{Exclusion Distance Determination}
\label{sec:ed}

\subsubsection{Intuition.} 
\label{sec:intuition}
Although repelling NTD a sufficient distance would easily make TD closer to $\bm{q}$, the HNSW traversal mechanism \zl{still} needs neighbors to \zl{maintain} graph connectivity, \zl{ensuring the sufficient traversal in nearest neighbor search.}
Thus, the exclusion distance must be narrowly bounded within a delicate range.

\begin{figure}[t]
    \centering
    \includegraphics[width=0.65\linewidth]{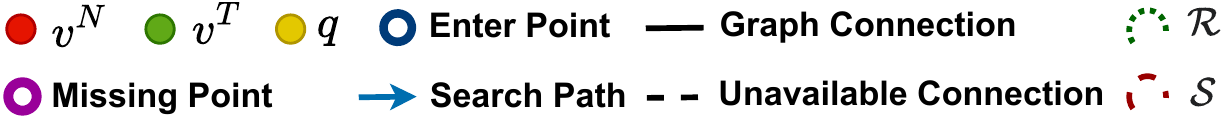}
    \subfloat[Original Distribution]{
        \includegraphics[width=0.25\linewidth]{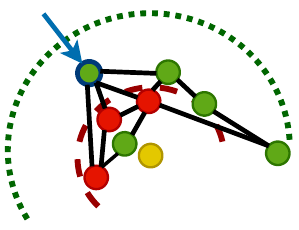}
        \label{fig:dis1}
    }
    \subfloat[Excessive $D$]{
        \includegraphics[width=0.25\linewidth]{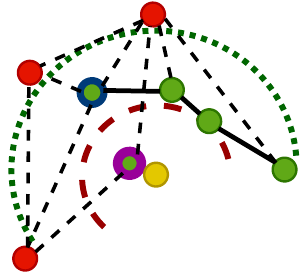}
        \label{fig:dis2}
    }
    \subfloat[Ideal $D$]{
        \includegraphics[width=0.25\linewidth]{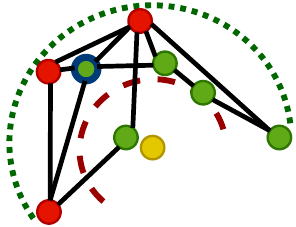}
        \label{fig:dis3}
    }
    \vspace{-2mm}
    \caption{The vector distance distributions under various $D$.}
	\label{fig:distance_1}
\end{figure}

As shown in Figure \ref{fig:dis1}, the search path directs to a new vector and $\mathcal{R}$ is full. This vector connects a $\bm{v}^{T}$ and three $\bm{v}^{N}$. If the current vector distance distribution remains unchanged or if $D$ is insufficient, the three NTD points will enter $\mathcal{R}$. Due to their extreme proximity to $\bm{q}$, they persistently occupy the positions in 
$\mathcal{R}$, preventing other promising TD points from entering. When three NTD points violate the condition shown in line 15 of Algorithm \ref{alg:optigreedy} due to excessive $D$, they cannot enter the $\mathcal{R}$. As shown in Figure \ref{fig:dis2}, their connections are unavailable. Consequently, graph traversal operations initiated from any direction cannot reach the TD point (purple circle) nearest to the $\bm{q}$ under this compromised graph connectivity. The ideal $D$ appears in Figure~\ref{fig:dis3}, where all NTD points remain confined within the region of $\mathcal{R}$ but lie well beyond the radius of $\mathcal{S}$. This vector distance distribution supports both the index connectivity and the accuracy of top-$k$ query results. To this end, an ideal $D$ for NTD should be in the following range:
\begin{equation}
\label{eq:distance}
    \max_{\bm{v}^T \in \mathcal{S}} Dis(\bm{q}, \bm{v}^T) < Dis(\bm{q},\bm{v}^{N})+D < \max_{\bm{v} \in \mathcal{R}} Dis(\bm{q}, \bm{v}),
\end{equation}
where $\max_{\bm{v}^T \in \mathcal{S}} Dis(\bm{q}, \bm{v}^T)$ is the farthest $\bm{v}^{T}$ from $\bm{q}$ in $\mathcal{S}$, and $\max_{\bm{v} \in \mathcal{R}} Dis(\bm{q}, \bm{v})$ is the farthest $\bm{v}$ from $\bm{q}$ in $\mathcal{R}$. 
The $D$ in this range can balance ANNS efficiency with index connectivity. 

\subsubsection{Theoretical Derivation.}
\label{subsubsec:derivation}

Based on the above analysis, we derive the optimal value of $D$ by identifying the key parameters that significantly affect retrieval performance, including the cardinality of the candidate set $\mathit{ef}$, the cardinality of the result set $k$, the selectivity $p$, and the average distance between data points $\Delta d$.
To systematically examine the effect of each parameter on $D$ and theoretically derive its optimal value, we adopt a control variate methodology and model the influence of unmeasured factors through a residual function $x$.
The theoretical derivation in this section assumes that the attribute distribution among data points has no correlation with their positions in the vector space. 

The derivation of $D$ begins with $\Delta d$, a parameter obtained from the dataset's vector distance distribution during indexing. 
This formulation is theoretically grounded in a classical spatial statistical model that assumes data points are independently and uniformly distributed in a $d$-dimensional Euclidean space, loosely corresponding to a homogeneous Poisson point process \cite{Poisson}. 
Under this idealized model, the expected distance $D_m$ from an arbitrary point to its $m$-th nearest neighbor satisfies the asymptotic relationship $ d_m \propto m^{1/d} $. 
To achieve a feasible approximation, we perform a Taylor expansion of the function  $f(m) = m^{1/d}$ around a reference point $m_0$, where $m_0$ is deemed as $ef$ (around $\mathit{ef} = 100$). The expansion is given by
\begin{equation}
f(m)=m_0^{1/d}+\frac{1}{d}m_0^{\frac{1}{d}-1}(m-m_0)+\frac{1}{2d}m_0^{\frac{1}{d}-2}(m-m_0)^2+\dots.
\end{equation}
Given that $d > 1$ and $m_0$ is sufficiently large, the higher-order Taylor coefficients (proportional to $m_0^{1/d - n}$ for $n \geq 2$) are tiny compared to the zeroth- and first-order terms, which can be safely ignored in the local approximation.
Therefore, the relationship between $d_m$ and $m$ is well approximated by a linear function within a neighborhood of $m_0$.
      
Although real-world spatial distributions often deviate from an ideal Poisson process, the general rule still holds: nearest-neighbor distances increase monotonically as a function of $m$. We observe that the linear fitting strategy can model the spatial distribution with acceptable empirical error. This practical operation enables and grounds the robust estimation of local density $\Delta d$ through a linear model derived from the distance distribution. Theoretically, since different queries induce distinct $\Delta d$, they could benefit from customized $D$ values for optimal retrieval performance.
However, in practice, determining an optimal $D$ for a query requires counting $\Delta d$, which incurs prohibitively high costs in real-time query serving. To address this issue, we determine $D$ with a strategy using the dataset's global $\Delta d$, which is calculated during the offline index construction phase. That avoids computing the local density $\Delta d$ online. The global $\Delta d$ is defined as

\begin{equation}
    \label{eq:avedistance}
    \Delta d = \frac{d_\alpha - d_\beta}{\alpha-\beta},
\end{equation}
where $d_\alpha$ and $d_\beta$ represent \zl{the} average distances to the $\alpha$-th and the $\beta$-th nearest vectors from a vector, respectively. Different from the calculation on averaging the distance to all the neighbors, our definition is practical for ANNS, as shown in Section \ref{subsubsec:linears}.

Due to the fact that a larger $\Delta d$ necessitates a correspondingly larger $D$ to ensure that an NTD point is pushed to a position equivalent to that of a TD point, $D \propto \Delta d$. We define $D$ as
\begin{equation}
\label{eq:xandd}
D = x \Delta d,
\end{equation}
where $x$ is a function related to $\mathit{ef}$, $k$, and $p$. 
Based on Equation (\ref{eq:avedistance}), when the $\bm{q}$ is deemed as the anchor vector, the distance between $\bm{q}$ and $m$-th nearest vector to $\bm{q}$ can be expressed by 
$$
d_{m} \approx d_0 + (m-1)\Delta d,
$$
where $d_0$ is $Dis(\bm{q},\bm{v}_0)$ and $\bm{v}_0$ is the nearest vector to the $\bm{q}$. Therefore, the coverage of a nearest vector set based on the $\bm{q}$ can be expressed by $d_m$, where $\bm{v}_m$ is the $m$-th nearest vector to the $\bm{q}$. Given $\mathcal{S}$ and $\mathcal{R}$ contain $k$ and $\mathit{ef}$ vectors, respectively,
the radii of $\mathcal{S}$ and $\mathcal{R}$ for the $\bm{q}$ shown in Figure~\ref{fig:distance_1} can be expressed by 
\begin{numcases}{}
R(\bm{q},\mathcal{S}) = d_0 + (k-1)\Delta d, \label{eq:rdiss} \\
R(\bm{q},\mathcal{R}) = d_0 + (\mathit{ef}-1)\Delta d, \label{eq:rdisr}
\end{numcases}
where $R(\cdot)$ denotes the radius.

To meet the
left part of Inequality (\ref{eq:distance}) 
based on Equation (\ref{eq:rdiss}), all NTD points in $\mathcal{S}$ must be excluded from $\mathcal{S}$ after the $D$ is applied. In the extreme case where the $\bm{v}_0$ is an NTD point, the $\inf D = R(\bm{q},\mathcal{S})$. Meanwhile, to satisfy $|\mathcal{S}|=k$, $R(\bm{q},\mathcal{S})$ must expand to cover the top $\frac{k}{p}$ nearest TD before the vector distance distribution changes. This process aims to find the support vector from the $\bm{v}_k$ to the $\bm{v}_{\frac{k}{p}}$ based on an expansion factor of $\frac{1}{p}$. Similar to
the right part of Inequality (\ref{eq:distance}) 
based on Equation (\ref{eq:rdisr}), the $R(\bm{q},\mathcal{R})$ also expands after the $D$ is applied. Consequently, the support vector for $\mathcal{R}$ alters to the $\bm{v}_{\frac{\mathit{ef}}{p}}$. As a result, the radii of $\mathcal{R}$ and $\mathcal{S}$ under the new vector distance distribution become
\begin{numcases}
{}
R(\bm{q},\mathcal{S}) = d_0 + (\frac{k}{p}-1)\Delta d, \label{eq:nrdiss} \\
R(\bm{q},\mathcal{R}) = d_0 + (\frac{\mathit{ef}}{p}-1)\Delta d. \label{eq:nrdisr}
\end{numcases}

In addition, the $D$ is applied only to NTD points that constitute a proportion of ($1-p$) of $\mathcal{D}$. Therefore, the additional average distance assigned to each $\bm{v}^{N}$ and $\bm{v}^{T}$ are $\frac{D}{1-p}$ and $\frac{pD}{1-p}$, respectively. It causes the $R$ in Equation (\ref{eq:nrdisr}) and Equation (\ref{eq:nrdiss}) to be systematically inflated. To ensure accuracy, the calculation of $R(\bm{q},\mathcal{S})$ and $R(\bm{q},\mathcal{R})$  incorporates an offset of $\frac{pD}{1-p}$ applied to the global average distance.

Based on the above analysis, Inequality (\ref{eq:distance}) can be derived to the form as
\begin{numcases}{}
d_0+(\frac{k}{p}-1) \Delta d + D < d_0 + (\frac{\mathit{ef}}{p}-1) \Delta d - \frac{Dp}{1-p}, \label{eq:1} \\
d_0 + D > d_0 + (\frac{k}{p}-1) \Delta d - \frac{Dp}{1-p}, \label{eq:2}
\end{numcases}
where Inequality (\ref{eq:1}) ensures that the NTD point in $\mathcal{S}$ with the farthest distance to $\bm{q}$ remains in $\mathcal{R}$ even when its distance is augmented by $D$, while Inequality (\ref{eq:2}) ensures that the NTD point in $\mathcal{S}$ with the nearest distance to $\bm{q}$ is excluded from $\mathcal{S}$ after augmentation. Through the derivation of two inequalities, we establish that
\begin{equation}
\label{eq:disres}
(1-p)(\frac{k}{p}-1)\Delta d < D < (1-p)(\frac{\mathit{ef}}{p} - \frac{k}{p})\Delta d.
\end{equation}
According to the minimax principle in optimization theory, selecting the average of the two bounds provided by the inequalities yields the most robust estimator. Therefore, we strategically recommend the intermediate value for $D$. It is described as
\begin{equation}
\label{eq:half}
D = \frac{1}{2p} (1-p)(\mathit{ef} - p) \Delta d.
\end{equation}

Empirically, normalizing this value by $ef$ is found to further enhance robustness across different search parameters. Equation (\ref{eq:half}) demonstrates that $D$ is sensitive to $p$ and their relationship is consistent with intuitive reasoning. For example, a negative correlation exists between $p$ and $D$. When $p\to 0$, $D \to \infty$, while as $p\to 1$, $D \to 0$.

\subsection{Termination Condition Optimization}
\label{sec:tc}
The conventional termination condition of HNSW \zl{is satisfied when} $\min_{\bm{v} \in \mathcal{C}} Dis(\bm{q}, \bm{v}) > \max_{\bm{v} \in \mathcal{R}} Dis(\bm{q}, \bm{v})$. 
Since the vector distance distribution has changed by $D$ in our approach, we adopt a similar condition \zl{to terminate the query search, \ie,}
$$\min_{\bm{v} \in \mathcal{C}} \overline{Dis}(\bm{q}, \bm{v}) > \max\limits_{\bm{v} \in \mathcal{R}} \overline{Dis}(\bm{q}, \bm{v}).$$ 
However, such a condition could introduce the issue of premature termination.

\subsubsection{Issue of Premature Termination.}

The conventional termination condition effectively balances search efficiency and recall under the implicit assumption that all candidate vectors are TD points.
In the filtered ANNS setting, this assumption no longer holds.
Even when the termination condition is satisfied, further exploration may still reveal TD points that are closer to $\bm{q}$.
If a sequence of NTD points is consecutively added to both $\mathcal{C}$ and $\mathcal{R}$ during traversal, the decision to terminate will be dominated by the distances among these NTD points.
Since the $\Delta d$ between NTD points remains unchanged, the termination condition may trigger prematurely once no nearer NTD points are found while some TD points remain undiscovered.
This premature termination leads to an incomplete search and, consequently, degraded recall.

\subsubsection{Optimization.}

To mitigate premature termination caused by consecutive NTD traversals, we adopt a conservative search strategy.
This strategy aims not only to obtain $k$ TD points in the final $\mathcal{S}$, but also to ensure that sufficient TD points are included in the $\mathcal{R}$ to make termination decisions reliable.
Let $\overline{p}$ denote the proportion of TD points within $\mathcal{R}$.
To guarantee that at least $k$ TD points can be retrieved, the candidate distribution must satisfy $\mathit{ef}\cdot\overline{p}-k>0$, which can be rewritten as
\begin{equation}
\overline{p} > \frac{k}{\mathit{ef}}.
\end{equation}
Combining Equation~(\ref{eq:1}) and Equation~(\ref{eq:2}), this constraint is satisfied only when $\mathit{ef} > 2k$, which ensures $\overline{p} < 0.5$.
However, increasing $\overline{p}$ generally leads to higher computational overhead due to longer search paths.
To balance efficiency and recall, we empirically set $\overline{p}=0.5$, requiring that at least half of the points in $\mathcal{R}$ are TD points.
Finally, we refine the original termination condition by incorporating this additional constraint to form a more robust termination criterion.
\ourmethod treats this optimization as an optional component for high-precision search. When activated, it can effectively improve the recall rate while keeping the $\mathit{ef}$ parameter unchanged.

%% file: evaluation.tex
\section{EVALUATION}
\label{sec:evaluation}
We comprehensively evaluate the performance of \ourmethod across multiple datasets and various filtering scenarios against state-of-the-art methods. 
Specifically, Section \ref{subsec:setup} describes the experimental setup. Section \ref{subsec:perf} evaluates search performance across different scenarios. Section \ref{subsec:construction} analyzes the indexing construction overhead, Section \ref{sec:6_4} presents an ablation study to evaluate the effectiveness of each component, and Section \ref{sec:verifi} validates the correctness of our theoretical assumptions. All experiments are conducted on a server equipped with two Intel(R) Xeon(R) Gold 6326 CPUs @ 2.90GHz (32 cores and 64 threads in total) and 512 GB of memory.

\subsection{Experimental Setup}
\label{subsec:setup}

\subsubsection{Filtering Conditions.} 
We employ four types of filtering conditions as experimental scenarios. 
The specific settings are as follows:
\begin{itemize}[noitemsep,topsep=0pt,parsep=0pt,partopsep=0pt,leftmargin=*]
    \item \textbf{Equality:} 
    The attribute value must exactly match the queried value.
    We consider two equality conditions: (1) a boolean equality condition, \textit{Equality\_bool}, with a selectivity of 50\%; and (2) an integer equality condition, \textit{Equality\_int}, with a selectivity of 10\%.
    
    \item \textbf{Inclusion:} The value of an attribute belongs to the set specified by the filtering condition. We use a set of three values for an integer attribute, corresponding to a selectivity of 30\%. 
    
    \item \textbf{Range:} The value of an attribute lies in a given value range. 
    We draw a range with 10\% and 50\% selectivity as \textit{Range\_10} and \textit{Range\_50}. 
    
    \item \textbf{Logic:} It combines the above filtering conditions using logical operators (AND, OR, and NOT) to form a composite filtering condition. 
    \zl{
    In our experiments, we evaluate a logical AND combination of \textit{Equality\_int} and \textit{Range\_50}, yielding an overall selectivity of approximately 5\%.
    }
\end{itemize}

\subsubsection{Datasets.} 
Table \ref{tab:dataset} presents all datasets used in our evaluation.
For each dataset, we \zl{attach} the vectors with attributes to support diverse testing scenarios. 
Each vector is assigned three types of attributes: \textit{bool}, \textit{int}, and \textit{float}. 
For \textit{bool} attributes, true and false are assigned with equal probability. 
For \textit{int} attributes, values are uniformly sampled from the integers 0 to 9. 
For \textit{float} attributes, values are uniformly drawn from the range [0, 100].

\begin{table}[t]
    \caption{Dataset statistics.}
    \vspace{-3mm}
    \label{tab:dataset}
    \centering
    \scalebox{0.9}{
    \begin{tabular}{c|c|c|c|c}
    \hline
        \toprule
        \textbf{Dataset} & \textbf{\# Base} & \textbf{\# Query} & \textbf{Dimension} & \textbf{Source} \\
        \midrule
        \textbf{SIFT1M} & 1000000 & 10000 & 128 & Image \\ \hline
        \textbf{GIST1M} & 1000000 & 1000 & 960 & Image \\ \hline
        \textbf{DEEP1M} & 1000000 & 10000 & 96 & Image \\ \hline
        \textbf{Msong} & 992272 & 200 & 420 & Audio \\ \hline
        \textbf{Paper} & 2029997 & 10000 & 200 & Text \\ \hline
        \textbf{Words} & 8000   & 200 & 3072  & Text \\ \hline
        \textbf{LAION25M} & 25011200 & 1000 & 512 & Text$\And$Image \\
        \bottomrule
    \end{tabular}
    }
\end{table}

\subsubsection{Baselines.}
We compare our approach against six state-of-the-art filtered ANNS methods and two vector databases. We use their open-sourced codes unless otherwise noted. We configure the vector databases to use HNSW. Below, we describe the configuration of each baseline.
\begin{itemize}[noitemsep,topsep=0pt,parsep=0pt,partopsep=0pt,leftmargin=*]
    \item \textbf{ACORN-$\gamma$}: We follow the original paper~\cite{ACORN}, setting $M=32$ and $M_\beta = 64$. $\gamma$ is set to 12 for all datasets but GIST1M. Due to its higher dimensionality, we set $\gamma = 50$ for GIST1M.
    \item \textbf{ACORN-1}: We set $ M = M_\beta = 32 $ and $ \gamma = 1 $ for all datasets, according to~\cite{ACORN}.
    \item \textbf{Result Set Filtering (RSF)}: We implement this baseline on \texttt{hnswlib}\footnote{https://github.com/nmslib/hnswlib}. The HNSW index is constructed with $ M = 32 $ and $\mathit{efc} = 40$ for \texttt{SIFT1M}, \texttt{DEEP1M}, \texttt{Paper}, and \texttt{Words}, and with $ M = 64 $ and $\mathit{efc} = 200$ for \texttt{GIST1M}, \texttt{Msong} and \texttt{LAION25M}.
    \item \textbf{Milvus}~\cite{milvus}: As a widely used vector database supporting multiple indexing algorithms, we configure Milvus to use HNSW as the underlying index. We set the same $M$ and $\mathit{efc}$ as in the Result Set Filtering for the fair comparison.
    \item \textbf{Vearch}~\cite{vearch}: It is a distributed vector database developed by JD.com. We also configure it to use HNSW indexing with identical parameters to those used in Result-Set-Filtering.
    \item \textbf{UNG}: We implement UNG following the methodology in~\cite{ung}, which builds proximity graphs using the Vamana algorithm. The parameters are set as $ \alpha = 1.2 $, degree bound $ R = 32 $, and maximum queue length $ L = 100 $, cross-group edges $ \delta = 6 $ and the number of entry points is $ \sigma = 16 $. Note that UNG requires attribute values to start from 1. Therefore, we shift the original bool attributes (0/1) to the range [1, 2], and int attributes (0--9) to [3, 12]. Since UNG does not support range queries on float attributes, we omit float attributes during indexing. We adopt the containment scenario for implementing \textit{Equality} and the overlap scenario for \textit{Inclusion}.
    \item \textbf{SeRF}: 
    For range query evaluation, we follow the 2DSegmentGraph construction method in~\cite{SeRF}. The graph is built on top of HNSW. For \texttt{SIFT1M}, \texttt{DEEP1M}, \texttt{Paper} and \texttt{Words}, we set $ index_k = 32 $, $ \mathit{ef}\_con = 40 $; for the remaining datasets, we use $ index_k = 64 $, $ \mathit{ef}\_con = 200 $. For all datasets, we set $\mathit{efc}\_max=500$. SeRF only supports range filtering and does not support float attributes. Therefore, we convert float attributes into ordered integers and assign them as int attributes for SeRF.
\end{itemize}

We released our code on GitHub\footnote{\url{https://github.com/JunjieSong123/FAVOR}}.
\noindent{\ourmethod} is implemented in C++ and
we use the same HNSW construction parameters as the above baselines for fairness. 
Additionally, we set $ \alpha = 10 $ and $ \beta = \mathit{efc} $ for computing $ \Delta d $.
The search performance is evaluated under single-threaded execution with SIMD optimizations disabled since not all methods support multi-threading or SIMD.

\subsection{Performance Evaluation}
\label{subsec:perf}

\begin{figure*}[t]
  \centering
  \includegraphics[width=0.75\linewidth]{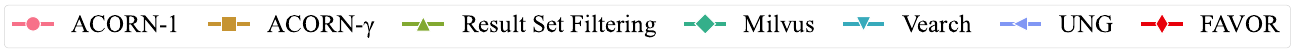}\vspace{0mm}\\
  \includegraphics[width=0.95\linewidth]{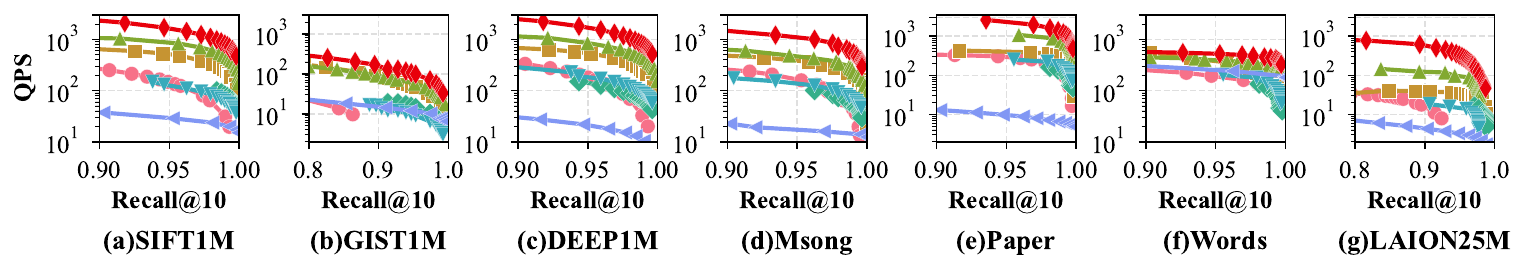}\vspace{0mm}\\
  \includegraphics[width=0.95\linewidth]{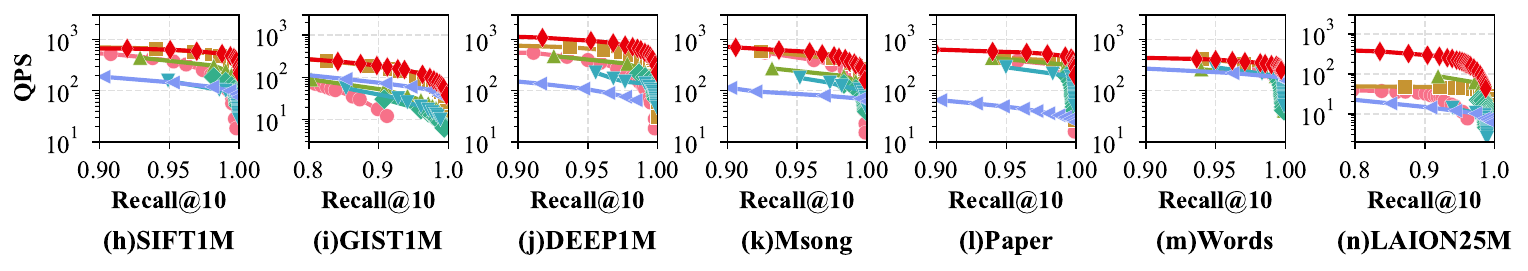}\vspace{0mm}

  \vspace{-0.4cm}
  \caption{Search performance in \textit{Equality\_bool} ( (a) -- (g) ) and \textit{Equality\_int} ( (h) -- (n) ).}
  \Description[equality condition]{equality-condition}
  \vspace{-0.4cm}
  \label{fig:eval:equality}
\end{figure*}

\begin{figure*}[t]
  \centering
  \includegraphics[width=0.98\textwidth]{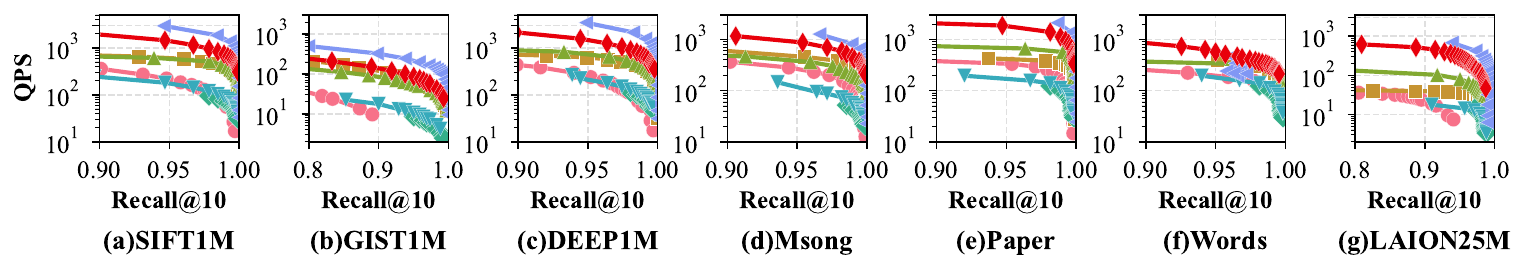}
  \caption{Search performance in \textit{Inclusion}.}
  \Description[inclusion condition]{inclusion-condition}
  \label{fig:eval:inclusion}
\end{figure*}

\subsubsection{QPS-recall tradeoff comparison}
\label{subsubsec:qps-recall}
\begin{figure*}[t]
  \centering
  \includegraphics[width=0.7\linewidth]{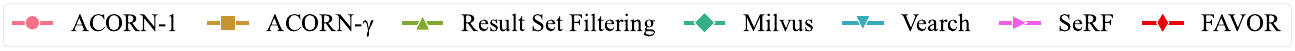}\vspace{0mm}\\
  \includegraphics[width=0.95\linewidth]{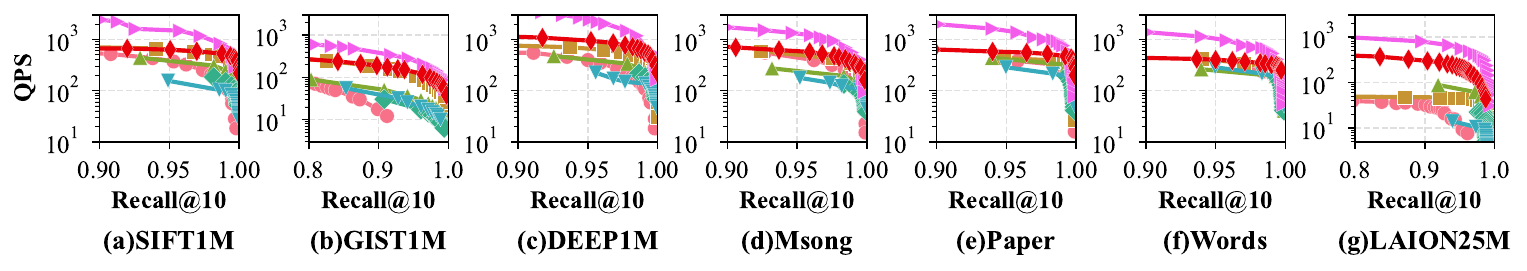}\vspace{0mm}\\
  \includegraphics[width=0.95\linewidth]{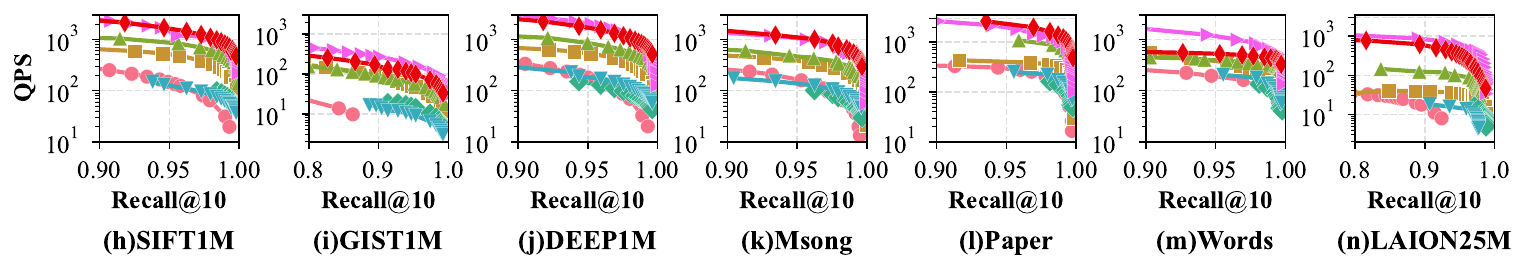}\vspace{0mm}
  \vspace{-0.4cm}
  \caption{Search performance in \textit{Range\_10} ( (a) -- (g) ) and \textit{Range\_50} ( (h) -- (n) ).}
  \Description[range condition]{range-condition}
  \label{fig:eval:range}
\end{figure*}

\begin{figure*}[t]
  \centering
  \includegraphics[width=0.7\linewidth]{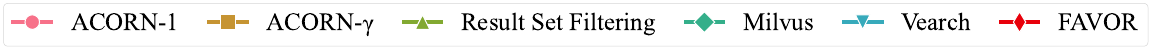}\vspace{0mm}\\
  \includegraphics[width=0.95\linewidth]{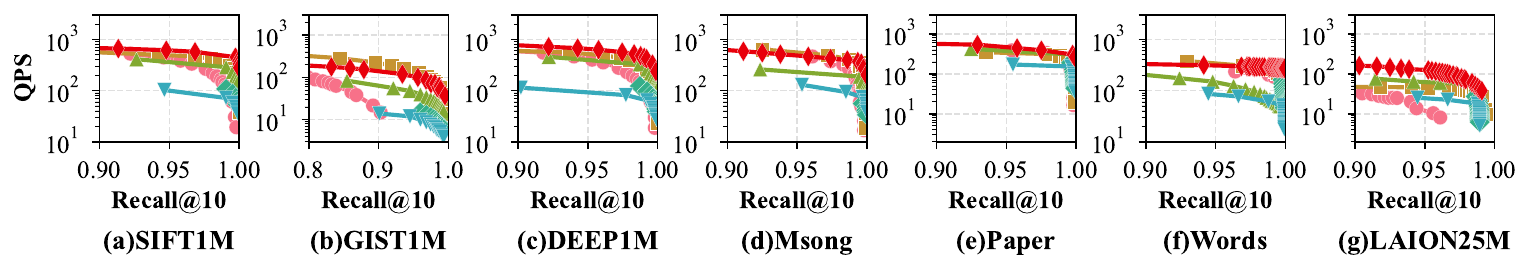}\vspace{0mm}
  \vspace{-0.4cm}
  \caption{Search performance in \textit{Logic}.}
  \Description[logic condition]{logic-condition}
  \label{fig:eval:logic}
\end{figure*}

Figure~\ref{fig:eval:equality}, ~\ref{fig:eval:inclusion}, ~\ref{fig:eval:range}, and ~\ref{fig:eval:logic} present the QPS-recall curve at high recall for the four filtering conditions.

As shown in Figure~\ref{fig:eval:equality}, in the \textit{Equality} scenario, \ourmethod outperforms all other baselines across all datasets. 
Under the condition of $Recall@10 = 95\%$, the performance improvement of \ourmethod over the best baseline ranges from 1.3$\times$ to 5$\times$.

Figure~\ref{fig:eval:inclusion} presents the results for the \textit{Inclusion} scenario. 
\ourmethod still outperforms all filtering-agnostic baselines, achieving a performance improvement of 1.5$\times$ to 3$\times$ when $Recall@10 = 95\%$. 
Note that the performance of \ourmethod is also competitive when compared to the filter-specific method UNG.

In the \textit{Range} scenario, as shown in Figure \ref{fig:eval:range}, \ourmethod outperforms all filtering-agnostic methods. Remarkably, \ourmethod achieves performance comparable to SeRF on the \texttt{SIFT}, \texttt{Msong}, \texttt{Paper}, and \texttt{LAION25M} datasets within the \textit{Range\_50} setting, despite SeRF’s specialized optimization for range queries.

In the \textit{Logic} scenario, as shown in Figure \ref{fig:eval:logic}, \ourmethod ranks among the top-performing methods, trailing only slightly behind ACORN-$\gamma$ on \texttt{GIST1M}.
This minor gap arises because ACORN-$\gamma$ (with $\gamma=50$) employs a densely connected index structure that incurs substantial additional indexing overhead (see Table \ref{tab:index_time}), yielding higher performance in this specific case.

Overall, across \zl{nearly all} experimental scenarios, \ourmethod consistently outperforms all filtering-agnostic baselines, including ACORN-1, ACORN-$\gamma$, and result-set filtering. 
At $Recall@10 = 95\%$, \ourmethod improves QPS by up to 5$\times$ over the best baseline. 
Although UNG and SeRF demonstrate QPS advantages in certain datasets under the \textit{Inclusion} and \textit{Range} conditions, respectively, these benefits arise from their specialized optimizations. 
In contrast, \ourmethod sustains robust performance across diverse query scenarios by effectively balancing filtering generality and search efficiency without requiring extensive index modification.
Notably, \ourmethod sustains consistently high performance across diverse scenarios while avoiding the structural overhead associated with index-enhanced approaches.

\subsubsection{Performance at different recall levels}
\label{subsubsec:recall}
\begin{figure}[htbp]
  \centering
  \includegraphics[width=0.6\columnwidth]{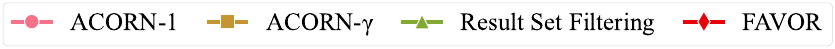}

  \begin{subfigure}[t]{\columnwidth}
    \centering
    \includegraphics[width=0.6\columnwidth]{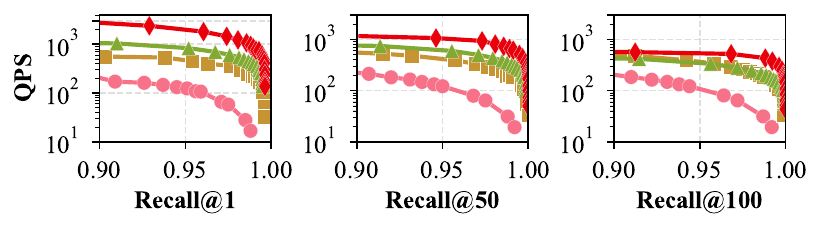}
    \caption{SIFT1M}
    \label{fig:recall_sift}
  \end{subfigure}
  
  \begin{subfigure}[t]{\columnwidth}
    \centering
    \includegraphics[width=0.6\columnwidth]{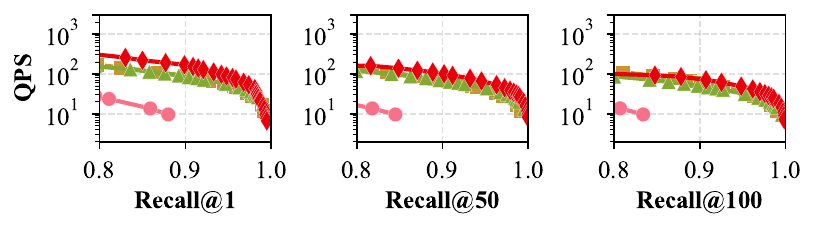}
    \caption{GIST1M}
    \label{fig:recall_gist}
  \end{subfigure}
  
  \caption{Search performance at different recall levels.}
  \label{fig:recall}
\end{figure}

We assess the proposed method in the \textit{equality\_bool} scenario across multiple recall levels. Experiments on \texttt{SIFT1M} and \texttt{GIST1M} datasets evaluate $Recall@1$, $Recall@50$, and $Recall@100$. As shown in Figure~\ref{fig:recall}, for $Recall>95\%$, \ourmethod substantially outperforms the baseline, achieving speedups of up to $3.5\times$ at $Recall@1$, $2.4\times$ at $Recall@50$, and $2.0\times$ at $Recall@100$. Performance gains are consistent across all recall levels and align closely with the trend observed at $Recall@10$.

\subsubsection{Performance under Varying Selectivity}
\label{subsubsec::selectivity}
\begin{figure}[ht]
    \centering
    \includegraphics[width=0.7\columnwidth]{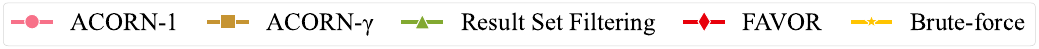} 
    \par\vspace{6pt}
    \vspace{-0.25cm}
    \begin{subfigure}[t]{0.35\columnwidth}
        \centering
        \includegraphics[width=\linewidth]{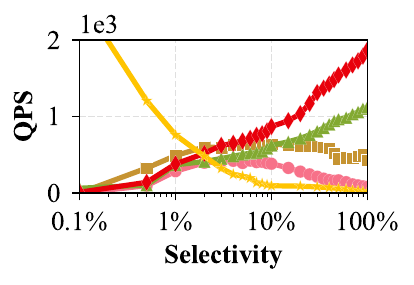}
        \vspace{-0.7cm}
        \caption{SIFT1M}
        \label{fig:fig:selectivity-sift}
    \end{subfigure}
    \begin{subfigure}[t]{0.35\columnwidth}
        \centering
        \includegraphics[width=\linewidth]{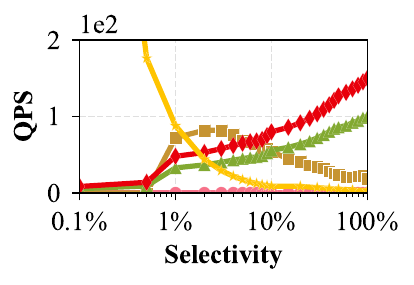}
        \vspace{-0.7cm}
        \caption{GIST1M}
        \label{fig:selectivity-gist}
    \end{subfigure}
    \vspace{-0.4cm}
    \caption{QPS yielded by different methods across varying selectivity at $Recall@10 = 95\%$.}
    \label{fig:selectivity}
\end{figure}

We assess \ourmethod's performance across different selectivity values to validate our search strategy. Under the \textit{Range} scenario with $Recall@10 = 95\%$, we compare the QPS of \ourmethod, including graph-based and brute-force search, against the baseline on \texttt{SIFT1M} and \texttt{GIST1M} datasets. Results in Figure~\ref{fig:selectivity} show how selectivity affects performance.

The two search strategies provided by \ourmethod, \ie, brute-force search and graph-based search, exhibit an interesting phenomenon across different experimental setups. Their performance curves consistently intersect within $1\% < p < 3\%$. When $p<1\%$, brute-force achieves higher QPS, and the QPS metric improves as $p$ decreases. Conversely, for $p>3\%$, the graph-based method delivers higher QPS, with performance improving as $p$ increases. Furthermore, within the intersection region, the QPS of the graph-based method varies by less than 8\%, while the brute-force method shows fluctuations exceeding 
50\%. 
This implies that the graph-based search is less sensitive to estimation error of $p$ than the brute-force search in this region. 

Based on this observation, we conservatively set the switching threshold at $p = 1\%$. At $p = 1\%$, the graph-based method enters its stable performance region, allowing the system to transition to a more robust search strategy upon reaching moderate selectivity levels. Even within the intersecting performance range, the QPS degradation remains below 8\% due to the flat response curve of the graph-based method. Therefore, the selected threshold is not only empirically supported but also strikes a balance between efficiency, stability, and robustness.

With the same settings, we also compare the performance of \ourmethod against baseline approaches.
Although the QPS decreases as $p$ lowers, \ourmethod mitigates the risk of sustained performance degradation by proactively employing brute-force search rather than graph-based search under low-selectivity conditions.
Therefore, \ourmethod consistently outperforms the state-of-the-art baseline methods across most selectivity levels. It achieves an average QPS improvement of $ 2.5\times $ on the \texttt{SIFT1M} dataset and $ 2\times $ on the higher-dimensional \texttt{GIST1M} dataset. These results demonstrate that \ourmethod remains effective and robust over a wide range of selectivity levels.

\subsection{Index Construction}
\label{subsec:construction}

\begin{table}[t]
\centering
\caption{Index construction time (s).}
\vspace{-2mm}
\scalebox{0.72}{
\begin{tabular}{c|c|c|c|c|c|c|c}
\toprule
\diagbox[width=10em]{Method}{Dataset} & \bf SIFT1M & \bf GIST1M & \bf DEEP1M & \bf Msong & \bf Paper & \bf {Words} & \bf {LAION25M}\\
\hline
\textbf{ACORN-1} & 7.2 & 31.8 & 5.5 & 17.2 & 22 & 0.8 & 697.9\\ \hline
\textbf{ACORN-$\gamma$}       & 150  & 3452  & 105   & 289    & 330   & 16.8  & 12791 \\ \hline
\textbf{Result Set Filtering} & 9.6  & 62.2 & 7.6  & 32.6      &  30.7 & 1.2 & 1147\\ \hline
\textbf{UNG}                  & 11.4     & 82.1     & 10.4     &  26.6  & 31.4  & 1.5 & 1216   \\ \hline
\textbf{SeRF}                 & 33.9     & 327      & 33.2     & 107.7     & 103.5  & 3.8 & 3782  \\ \hline
\textbf{\ourmethod}           & 11.2     & 64.7     & 9.7      & 34.1      & 32.5  & 1.3 & 1148 \\
\bottomrule
\end{tabular}
}
\label{tab:index_time}
\end{table}

\begin{table}[t]
\centering
\caption{Storage overhead for Index (GB).}
\vspace{-2mm}
\scalebox{0.72}{
\begin{tabular}{c|c|c|c|c|c|c|c}
\toprule
\diagbox[width=10em]{Method}{Dataset \& size}
& \makecell{\bf SIFT1M\\0.49} & \makecell{\bf GIST1M\\3.58} & \makecell{\bf DEEP1M\\0.37} & \makecell{\bf Msong\\1.56} & \makecell{\bf Paper\\1.52} & \makecell{\bf  Words\\0.094} & \makecell{\bf  LAION25M\\47.8}\\
\hline
\textbf{ACORN-1}      & 0.81  & 3.89 & 0.69 & 1.86 & 2.15 & 0.098 & 55.5\\ \hline
\textbf{ACORN-$\gamma$}      & 0.97      & 4.20     & 0.85     & 3.02     & 2.78  & 0.102 & 59.6  \\ \hline
\textbf{Result Set Filtering}        & 0.755     & 4.08     & 0.63     & 2.05     & 2.04 & 0.098 & 56.2      \\ \hline
\textbf{UNG}                         &  0.70  & 3.74 &  0.58 & 1.71 &  1.88 & 0.098 & 54.3  \\ \hline
\textbf{SeRF}                        & 0.78      & 4.33     & 0.69     & 2.13     & 2.36 & 0.10 & 57.4      \\ \hline
\textbf{FAVOR}                       & 0.755     & 4.08     & 0.63     & 2.05     & 2.04 & 0.098 & 56.2      \\

\bottomrule
\end{tabular}
}
\label{tab:index_size}
\end{table}

\subsubsection{Construction Time.} Table~\ref{tab:index_time} shows a comparison of index construction times for different methods under a 64-thread environment, where \ourmethod is built upon HNSW. During the construction process, we only utilize neighbor nodes within the $\mathit{efc}$ range as approximate $\alpha$-th and $\beta$-th nearest neighbors to estimate distances. 
This introduces only a minor additional cost when computing $\Delta d$, resulting in an average construction time about 2$s$ higher than that of Result Set Filtering, whose construction time is equivalent to vanilla HNSW.
This overhead is significantly lower than that of ACORN-$\gamma$ and SeRF, and is comparable to UNG. 
Considering \ourmethod’s superior adaptability and runtime performance, this modest overhead highlights its overall competitiveness.

\subsubsection{Storage Overhead.}
Table~\ref{tab:index_size} compares the index sizes of different methods with respect to the original dataset size. 
Noted that the reported index sizes include the original vectors. 
\ourmethod, built upon HNSW, only introduces \zl{small} additional storage overhead for attribute information and the data distribution parameter $\Delta d$, without incorporating any extra links or expanded neighbor lists. 
Consequently, its index size remains comparable to that of Result Set Filtering. 
Although the graph-tree hybrid index of UNG mitigates the overhead inherent to pure graph constructs, it, as well as SeRF, suffers from inflexibility in handling arbitrary predicates, necessitating compensatory storage. 
ACORN-$\gamma$, in contrast, depends on dense edge connections to maintain graph connectivity, resulting in substantial storage overhead. 
\ourmethod\ maintains connectivity without such redundancy, ensuring a more storage-efficient design.

\vspace{-4mm}
\subsection{Ablation Study}
\label{sec:6_4}

We conduct an ablation \zl{study} on the key components of \zl{our} \ourmethod search algorithm, including the exclusion distance mechanism and termination condition, to demonstrate their effectiveness. 

\begin{figure}
    \centering
    \includegraphics[width=0.65\linewidth]{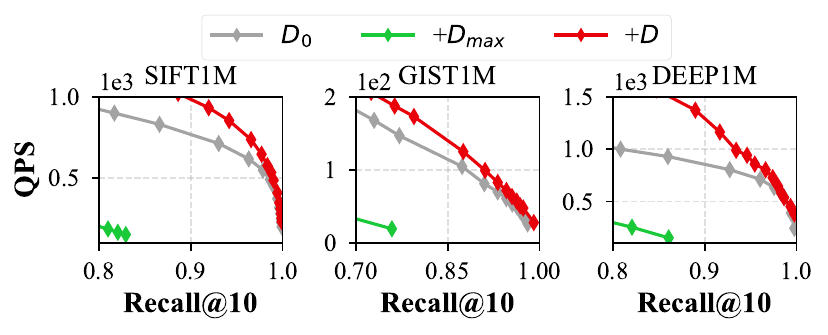}
    \vspace{-3mm}
    \caption{Effect of exclusion distance strategies.}
    \label{fig:distance}
\end{figure}

\subsubsection{Exclusion Distance} We conduct a comparative analysis of three \zl{exclusion} distances strategies using \textit{Equality\_int} on three datasets.
$D_0$ denotes setting the zero exclusion distance. 
The second strategy is 
$D_{max}=\max_{\bm{v}^{T}\in\mathcal{D}}Dis(\bm{q}, \bm{v}^{T})-\min_{\bm{v}^{N}\in\mathcal{D}}Dis(\bm{q}, \bm{v}^{N})$, as making all TD closer to the $\bm{q}$ than any NTD point.
$D$ is calculated by Equation (\ref{eq:half}) and used in \ourmethod.
Figure~\ref{fig:distance} shows that adopting $D$ achieves the best performance. 
Compared to the method with $D_0$, it achieves an average 1.6$\times$ improvement in QPS when $Recall@10 = 90\%$, demonstrating the effectiveness of using exclusion distance. Furthermore, adopting $D$ significantly outperforms the solution with $D_{max}$, which is consistent with the \zl{intuition} presented in Figure~\ref{sec:intuition}. 
Note that we observe that the advantage of employing $D$ is less pronounced \texttt{GIST1M}, because of its higher dimensionality and a relatively smaller $\Delta d$ among neighboring points. Consequently, $D$ has a less significant impact on the distance distribution, which caused the limited QPS improvement. 

\subsubsection{Termination Condition.}
\label{subsub:termination}
\begin{figure}[t]
    \centering
    \includegraphics[width=0.7\linewidth]{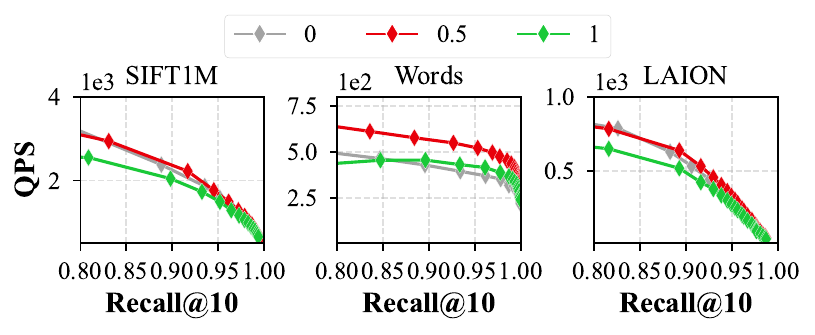}
    \vspace{-4mm}
    \caption{Impact of optimized termination condition.}
    \label{fig:termination}
\end{figure}

As analyzed in Section~\ref{sec:tc}, without optimizing the filtering condition, reliance solely on distance comparisons for search termination may lead to premature termination of the retrieval process.
In such cases, recall under the same search parameter $\mathit{ef}$ falls below the expected level, while QPS appears higher than anticipated.
To validate the selection of our termination condition parameter, we examine the trade-off between QPS and recall in the \textit{Equality\_bool} scenario across three datasets of varying scales: \texttt{SIFT1M}, \texttt{Words}, and \texttt{LAION25M}, using different values of the optimized termination threshold $\overline{p}$, where $\overline{p} = 0$ corresponds to the case without termination condition optimization.
As shown in Figure~\ref{fig:termination}, the optimal balance occurs at $\overline{p} = 0.5$, where the system achieves the highest QPS while still maintaining high recall rates ($Recall@10 > 95\%$).
Furthermore, by exploiting early termination, a smaller $\overline{p}$ can be further reduced to enable fast, lower-precision retrieval when speed is prioritized over accuracy.

\subsection{Verification Test}
\label{sec:verifi}

\subsubsection{Search path} 
\label{subsubsec:search_path}
\begin{figure}[t]
    \centering
    \includegraphics[width=0.7\linewidth]{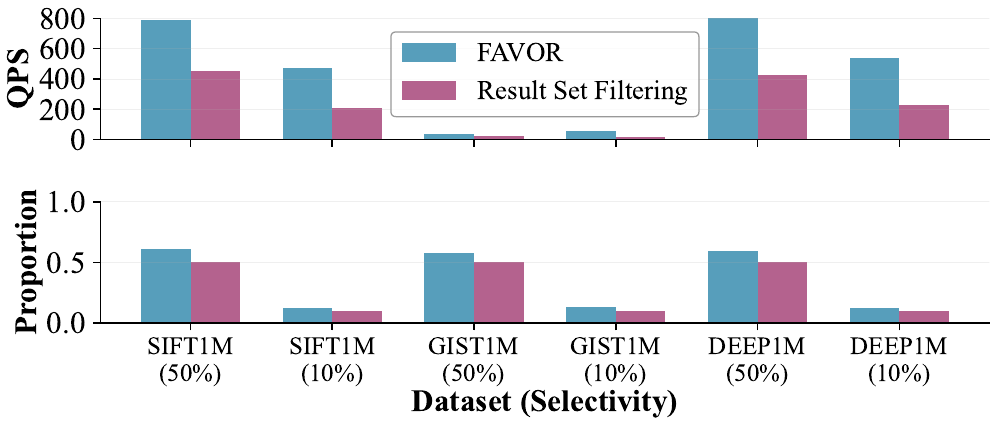}
    \vspace{-4mm}
    \caption{Performance vs. TD proportion on search paths.}
    \label{fig:tdrate}
\end{figure}

To validate the key idea presented in Section~\ref{sec:keyidea}, we conduct two sets of experiments on three datasets, including \texttt{SIFT1M}, \texttt{GIST1M}, and \texttt{DEEP1M}. First, we collect the proportion of TD points along the search paths generated by \ourmethod and Result Set Filtering. Figure~\ref{fig:tdrate} illustrates the relationship between QPS and the TD point proportion across different datasets and selectivity levels under the \textit{Equality} scenario at $Recall@10 = 95\%$. 
It is clear that there is a consistent positive correlation between the TD proportion and QPS for both methods.
Given that QPS is typically negatively correlated with search path length, this phenomenon implies that \ourmethod achieves shorter search paths under high-recall conditions, thereby indirectly validating the idea.

\begin{figure}[t]
    \centering
    \includegraphics[width=0.7\linewidth]{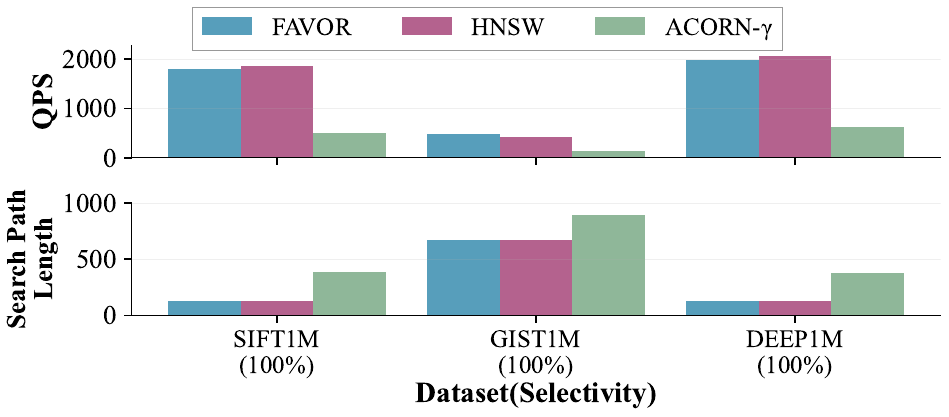}
    \vspace{-4mm}
    \caption{Comparison of search path length and QPS.}
    \label{fig:path}
\end{figure}

Second, under a non-filtering setting ($p = 100\%$) with $Recall@10 = 95\%$, we compare \ourmethod against vanilla HNSW and ACORN-$\gamma$ in terms of QPS and search path length.
As shown in Figure~\ref{fig:path}, \ourmethod achieves QPS and path lengths comparable to those of vanilla HNSW, and exhibits shorter paths and higher QPS compared to ACORN-$\gamma$. 
This phenomenon demonstrates that the exclusion distance mechanism maintains the graph connectivity, avoiding inefficient traversals.

\subsubsection{The Linear Model}
\label{subsubsec:linears}
We conduct experiments to validate the consistency between the proposed linear model and Equation (\ref{eq:avedistance}). Specifically, we examine whether the distance from a point in the space to its $m$-th nearest neighbor approximately follows a linear relationship with $m$. For each vector in the dataset $D$, we treat it as an anchor point and compute its distances to the $m$-th nearest neighbors, where $m = 1, 2, \dots, 1000$. We then perform linear regression with $m$ as the independent variable and the $d_m$ as the dependent variable.
The goodness of fit is assessed by the coefficient of determination, \ie, $R^2$\cite{Regress}. It is used to quantify the proportion of the variation in the dependent variable that is predictable from the independent variable. 
Its range is from 0 to 1, with higher values indicating a stronger linear relationship between the variables. 
As summarized in Table~\ref{tab:linear}, $R^2$ exceeds 0.8 across all datasets, with a consistently small standard deviation. 
Given high $R^2$ and low variance observed, for an arbitrary point in the space, the distance to its $m$-th nearest neighbor increases approximately linearly with $m$.
These results constitute strong empirical evidence supporting the validity of the proposed linear model.

\begin{table}[t]
\centering
\caption{Linear Model.}
\vspace{-2mm}
\resizebox{0.8\columnwidth}{!}{%
\begin{tabular}{c|c|c|c|c|c|c|c}
\toprule
\diagbox[width=8em, height=2em]{\textbf{Metric}}{\textbf{Dataset}} & 
\textbf{SIFT1M} & \textbf{GIST1M} & \textbf{DEEP1M} & \textbf{Msong} & 
\textbf{Paper} & \textbf{Words} & \textbf{LAION25M} \\
\midrule
\textbf{Mean $R^2$} & 0.848 & 0.835 & 0.856 & 0.864 & 0.867 & 0.817 & 0.821 \\
\hline
\textbf{$R^2$ Std.} & 0.035 & 0.046 & 0.044 & 0.037 & 0.048 & 0.061 & 0.083 \\
\bottomrule
\end{tabular}
}
\label{tab:linear}
\end{table}

%% file: relatedwork.tex
\section{Related Work}
\label{sec:related-work}


\textbf{ANNS.} ANNS techniques can be broadly classified into hash-based and graph-based methods.
Hash-based approaches~\cite{ANNS6} perform coarse-grained partitioning of the search space, struggling to achieve high recall efficiently for high-dimensional datasets.
Graph-based methods~\cite{ANNS1, ANNS2, ANNS3, ANNS7, ANNS8} explicitly encode accurate neighbor relationships into the indexing structure and have recently emerged as the dominant paradigm. \ourmethod builds on the success of the widely used HNSW graph and significantly improves its applicability in filtered ANNS.
Quantization~\cite{ANNS4, ANNS5, hakes} can be added to both hash-based and graph-based methods to reduce storage and speed up vector comparison by approximation with lossy representations, which can be added on top of \ourmethod to boost efficiency.

\noindent\textbf{Filtered ANNS.} Pre-filtering and post-filtering policies are common options in vector databases~\cite{milvus,vearch,AnalyticDB} due to their simplicity and versatility; however, the former is only efficient for low selectivity, and the latter often requires iterative search to return enough target points. Recent studies integrate filtering support into vector indexes to improve efficiency~\cite{Chronis25, NaviX}. SeRF~\cite{SeRF}, UNG~\cite{ung}, RangePQ~\cite{FANNS1}, DSG~\cite{FANNS2} are tailored methods for specific filtering conditions to achieve efficient search by additional index structures, which limit their flexibility and incur construction and maintenance overheads. \ourmethod only uses an exclusion distance during serving and does not modify the graph index structure, adding minimal overhead and offering competitive performance. SIEVE~\cite{SIEVE} supports arbitrary filters but groups vectors and maintains collections of subindexes, which means it still requires prior knowledge about queries. \ourmethod keeps a unified index and does not assume queries. ACORN~\cite{ACORN} searches only along TD points but suffers from connectivity issues, resulting in difficulty in achieving high recall under low selectivity. \ourmethod strikes a balance between prioritizing search on TD points and exploration with near NTDs, which improves the selectivity range suitable for graph-based search.

%% file: conclusion.tex
\section{Conclusion}
\label{sec:conclusion}

In this paper, we presented \ourmethod, a filter-agnostic vector ANNS method that delivers high efficiency across varying selectivity levels. It employs a selectivity-driven search selector to dynamically choose the more suitable approach between \ourmethod search algorithm and pre-filtering brute-force search to achieve higher QPS. \ourmethod search introduces a novel selectivity-aware exclusion distance mechanism for efficient and seamless integration with HNSW. \ourmethod achieves 1.3$\times$ to 5$\times$ higher QPS at $Recall@10 = 95\%$ than state-of-the-art filter-agnostic methods, while remaining competitive with tailored solutions in their target filtering scenarios. \ourmethod bridges the gap between generality and performance, making it suitable for a wide range of real-world applications.

%% file: Acknowledgments.tex
\begin{acks}
This work was achieved in Key Laboratory of Information Storage System and Ministry of Education of China. This work was supported by the National Key Research and Development Program Grant No.2023YFB4502701, the National Natural Science Foundation of China Grant No.62232007. This work was supported by the China Scholarship Council (No. 202406160132). 
\end{acks}